\documentclass[sigconf]{acmart} 
\AtBeginDocument{%
  \providecommand\BibTeX{{%
    \normalfont B\kern-0.5em{\scshape i\kern-0.25em b}\kern-0.8em\TeX}}}

\setcopyright{rightsretained}
\copyrightyear{2023}
\acmYear{2023}
\acmDOI{10.1145/3544548.3581569}

\acmConference[CHI '23]{Proceedings of the 2023 CHI Conference on Human Factors in Computing Systems}{April 23--28, 2023}{Hamburg, Germany}
\acmBooktitle{Proceedings of the 2023 CHI Conference on Human Factors in Computing Systems (CHI '23), April 23--28, 2023, Hamburg, Germany}
\acmPrice{15.00}
\acmISBN{978-1-4503-9421-5/23/04}

\usepackage{graphicx}
\usepackage{pythonhighlight}
\usepackage{caption}
\usepackage{subcaption}
\usepackage{csquotes}
\usepackage{framed}
\usepackage{acmart-taps}

\newsavebox{\measurebox}

\begin{document}

\title{Designing and Evaluating Interfaces that Highlight News Coverage Diversity Using Discord Questions}

\author{Philippe Laban}
\email{plaban@salesforce.com}
\affiliation{
  \institution{Salesforce AI Research}
  \country{United States}
}
\author{Chien-Sheng Wu}
\affiliation{
  \institution{Salesforce AI Research}
  \country{United States}
}
\author{Lidiya Murakhovs'ka}
\affiliation{
  \institution{Salesforce AI Research}
  \country{United States}
}
\author{Xiang `Anthony' Chen}
\affiliation{
  \institution{UCLA}
  \country{United States}
}
\author{Caiming Xiong}
\affiliation{
  \institution{Salesforce AI Research}
  \country{United States}
}



\begin{abstract}
Modern news aggregators do the hard work of organizing a large news stream, creating collections for a given news story with tens of source options. This paper shows that navigating large source collections for a news story can be challenging without further guidance.
In this work, we design three interfaces -- the Annotated Article, the Recomposed Article, and the Question Grid -- aimed at accompanying news readers in discovering coverage diversity while they read. A first usability study with 10 journalism experts confirms the designed interfaces all reveal coverage diversity and determine each interface's potential use cases and audiences. In a second usability study, we developed and implemented a reading exercise with 95 novice news readers to measure exposure to coverage diversity. Results show that Annotated Article users are able to answer questions 34\% more completely than with two existing interfaces while finding the interface equally easy to use.
\end{abstract}

\maketitle

\section{Introduction}
\label{sec:introduction}

\begin{figure*}
    \centering
    \includegraphics[width=0.90\textwidth]{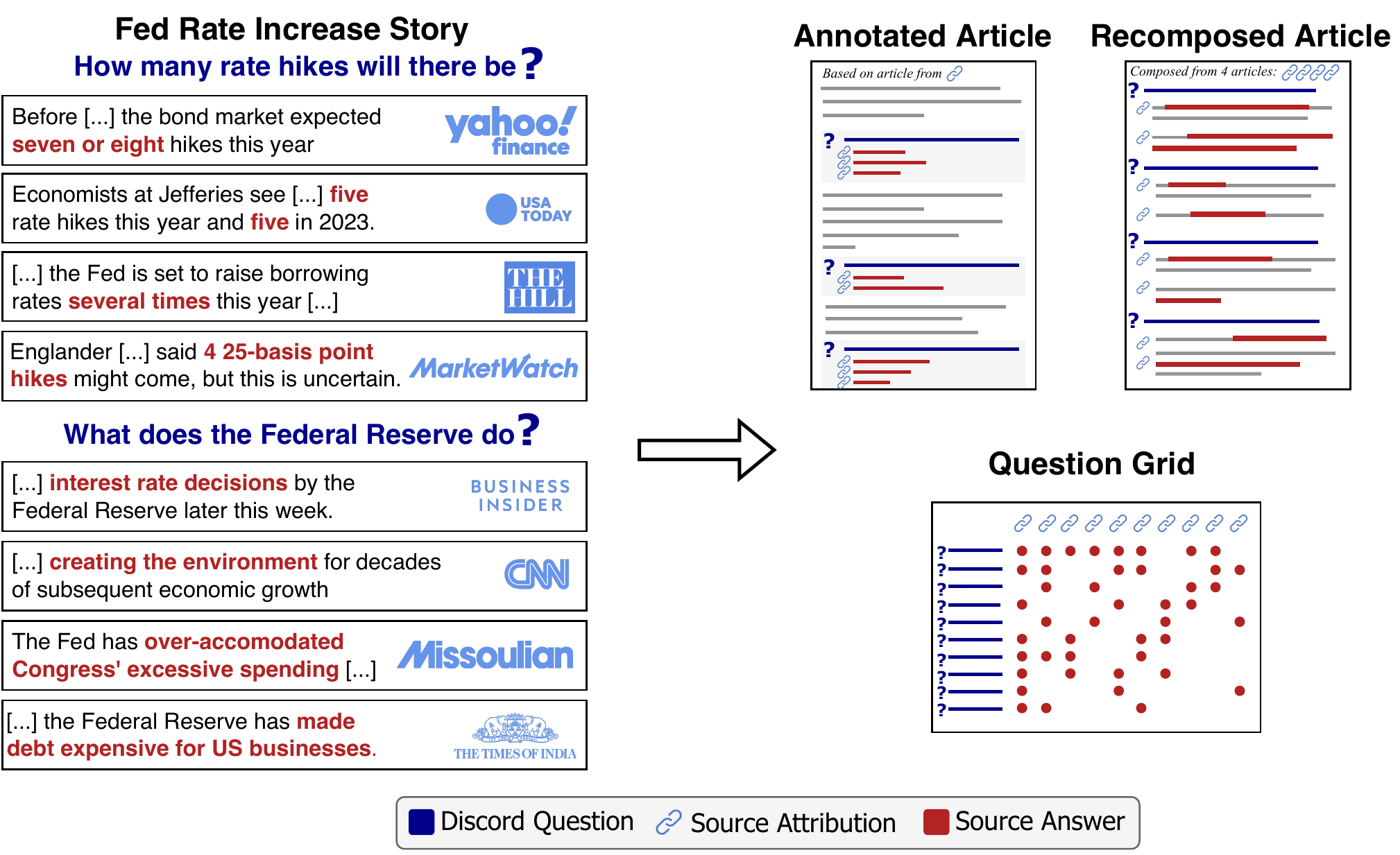}
    \caption{\textbf{We design and evaluate news reading interfaces that incorporate discord questions to reveal coverage diversity.} The Annotated Article is based on a single source article annotated with questions and answers from multiple sources. The Recomposed Article is a QA-format article composed from the ground up using multiple source articles. The Question Grid presents content on a \texttt{(question,source)}-matrix, specifying the answer each source provides to a list of questions.}
    \label{fig:high_level_examples}
\end{figure*}

In order to complement their understanding of political and world affairs, readers increasingly access the news through multiple channels, with 64\% of Americans relying on articles shared on social media \cite{rosenstiel_media_2014}, and 62\% using smart speakers to listen to the news \cite{smart_audio_report_2020}. In this multi-channel setting, it becomes important for news readers to have easy ways to compare and contrast opinions of varying sources \cite{perrault2019effects}, as a lack of transparency risks reader exposure to bias on critical societal issues such as elections or international affairs \cite{bernhardt2008political}.

Modern news aggregators play a crucial role in giving readers access to broad coverage diversity on any given topic, with features like ``See Full Coverage'' on Google News that put thousands of sources at the fingertips of news readers. In practice, however, news aggregator users must invest more time and effort to obtain broad coverage for a news story \cite{lee2015rise}, reading through several sources and sifting through overlapping content to build a more complete story understanding.

Some news aggregators provide additional guidance to simplify the source selection process, such as the political alignment of a source or information on the organization that owns each source \cite{allsides2021allsides}, assisting readers in anticipating potential bias. However, news aggregators treat news articles as \textit{atomic units} and do not typically help users compare details in coverage differences within news articles. The burden of aligning article text to compare source differences on story-specific issues is therefore left to the reader.

In this work, we explore an extended role of the news aggregator, in which details in coverage that sources diverge on are explicitly revealed, accompanying the reader in understanding the subtlety of the story. Elements of disagreement between sources are extracted using the Discord Questions framework \cite{laban2022discord}, an automatic pipeline based on Natural Language Processing (NLP).
For multi-source news stories, the framework defines discord questions as questions answered by a large proportion of the sources, with a level of diversity or disagreement in the sources' answers. Authors of the framework hypothesize that discord questions can be a powerful tool in helping news readers navigate source differences, with each question revealing how the sources side on a specific issue within the story.

As an illustration, Figure~\ref{fig:high_level_examples} gives an example of two generated discord questions for a story on inflation in Summer 2022 in the US. The questions highlight source disagreements on the role of the Federal Reserve and predictions of its future actions, revealing elements of the story-specific debate. Although promising, it is not evident how to integrate Discord Questions into news reading interfaces that can accompany novice users in realistic reading settings. In this work, we design three interfaces that leverage discord questions -- Annotated Article, Recomposed Article, and Question Grid -- and evaluate them in two usability studies. The three interfaces we propose vary in levels of complexity. With the Annotated Article, we select an existing news article and augment its contents with discord questions and selected multi-source answers. In the Recomposed Article, a QA-format article is \nobreak algorithmically \nobreak composed using the content of multiple sources. Finally, the \nobreak Question Grid achieves higher information density by laying out information within a \texttt{(question,source}-grid, specifying the answer each source provides to a list of questions.

We conduct the first usability study with 10 journalism experts. We ask participants to compare the proposed interfaces with two baselines (i.e., a plain News Article and a Headline List), with an aim to narrow down use cases and target audiences for each interface. Experts find that only the two baseline interfaces and the Annotated Article are suitable for novice news readers, with the Annotated Article giving the most complete overview of the story. The Question Grid is found to highlight the most coverage diversity overall and be relevant for advanced use cases such as helping newsrooms find unique perspectives to differentiate from the competition.

In a second usability study, we design a news-reading exercise to measure exposure to coverage diversity in a realistic news-reading scenario for novice readers, which we define as readers with no professional experience in journalism. We implement the exercise with 95 novice users who complete the exercise using three interfaces: the two baseline interfaces, and the Annotated Article. We find that the Annotated Article leads to a significantly broader understanding of news stories while being preferred in terms of ease of use. Our work makes the following contributions:
\begin{itemize}
    \item The design and implementation of three news interfaces\footnote{A live demonstration of the interfaces is available at \url{http://assembly.salesforceresearch.ai/}} -- the Annotated Article, the Recomposed Article, and the Question Grid -- based on the Discord Questions framework, aimed at highlighting news coverage diversity in multi-source news stories,
    \item The design of a news-reading exercise that measures reader exposure to news coverage diversity, in which readers are tasked to answer open-ended questions as thoroughly as possible, followed by manual scoring of answer completeness,
    \item An evaluation through two usability studies of proposed designs, confirming that integrating discord questions into news interfaces effectively highlights coverage diversity, and provides readers with a more complete story understanding in a realistic news reading situation, compared to existing baseline interfaces.
\end{itemize}

\section{Related Work}
\label{sec:related_work}

\subsection{Exposing Bias To News Readers.}

Prior work has proposed interfaces to accompany news readers in explicitly seeking diverse opinions. We classify approaches as supplying source indicators, annotating single articles, or multi-article interfaces.

\subsubsection{Source Position Indicators.} Source Position indicators typically depict the valence (left/right) and magnitude (moderate/extreme) of each source of information in a news interface \cite{liao2014can}. These indicators have become widespread, and are now integrated into popular aggregators such as the Wall Street Journal's Blue Feed Red Feed\footnote{https://graphics.wsj.com/blue-feed-red-feed/}, and AllSides \cite{allsides2021allsides}. Some studies have shown the effectiveness of source position indicators, for example, \citet{munson2013encouraging} built a system that sends a weekly report to readers summarizing the source positions they accessed, and showed that sending such reports increases reader visits to opposite or moderate news sites. One limitation of source indicators is their tendency to be overly generic, as source bias can vary based on the topic, and does not accompany the reader beyond article recommendation.

\subsubsection{Article Annotation.} Prior work has proposed annotations that augment and modify articles once they are selected by a reader. \citet{munson2010presenting} propose to highlight passages in articles that are in agreement with user opinion and to reorder article content such that agreeable content comes first, finding limited improvement in reader satisfaction. \citet{hamborg2020newsalyze} more directly go after educating readers, automatically highlighting terminology likely to be biased (e.g., freedom fighters vs. terrorists). Beyond annotating the content, prior work has also inserted information best practices in the form of a checklist to encourage readers to detect misinformation\cite{heuer2022comparative}. The single-source setting is limited, as the user cannot get exposed to content not present in the chosen source. We implement an interface in which we annotate a single article with contents from other sources, providing multi-source information within a single-source framing.

\subsubsection{Multi-Article Interface.} Some work has proposed interfaces to contrast and compare biases of multiple sources on a common story. Newscube\cite{park2009newscube, park2011newscube} proposes a clustering-based interface that organizes story sources by viewpoints. \citet{Hamborg2018BiasawareNA} propose NewsBird, a matrix-based news aggregator that organizes sources based on geopolitical origin, with a follow-up study showing the limited impact on user awareness of content bias\cite{Spinde2020EnablingNC}. Another approach to tackling news coverage diversity is to diversify the output during the user's search process, by modifying the recommender engine's results, for example, 3bij3\cite{loecherbach20203bij3} proposed a common framework to evaluate recommender systems, and \citet{heitz2022benefits} find that diversifying the recommendation of a news search engine can have depolarizing effects on news readers. Prior work most commonly treats individual articles as atomic units, limiting the scope to recommending a more diverse set of articles for a given topic to the user. 

In this work, we propose one article annotation interface, and two multi-article interfaces, all leveraging the common Discord Questions framework to extract annotations. Unlike prior work, the annotations are not solely focused on highlighting bias, but more generally the diversity of opinion, by bringing to light questions that receive diverse answers. 

\subsection{Surfacing Novel and Personalized Content.}

Another vein of work focuses on an information retrieval problem formulation, with an objective to assist a reader in discovering novel information, reducing a user's requirements to search through information surplus \cite{mcnair2006cultural}. \citet{iacobelli2010tell} build a system that recommends novel content to news readers, categorizing each recommendation as providing a novel quote, actor, or figure. NewsJunkie \cite{gabrilovich2004newsjunkie} leverages user history to tailor a news feed that prioritizes information novelty, while \citet{yom2014promoting} integrates information diversity directly into search engine results.

Other work has leveraged recent progress in automated question generation \cite{murakhovska2022mixqg} to personalize the presentation of news content, such as the question-driven news chatbot \cite{laban2020s}, or the NewsPod \cite{Laban2022NewsPodAA} project for automating Q\&A-based news podcasts.

In this work, we surface novel answers centered on questions for given stories and do not use personal user information or history to tailor the highlighted content.

\subsection{Surfacing Neutral Content.}

News bias can be argued to be detrimental to news readers, and some prior work has put forward methods to surface objective content and mute unwanted biases. \citet{babaei2018purple} leverage social media to find consensus articles, and create a ``Purple Feed'' that focuses on such articles that are well received by both sides of the political spectrum. Text generation, either through the form of full article generation \cite{lee2021mitigating} or multi-document summarization \cite{fabbri2019multi,Lee2022NeuSNM} has also been proposed as a method to mute bias in news. Prior work has however shown the negative effects of content moderation \cite{perrault2019effects}, finding that opinion heterogeneity provides users with a feeling of fairness. In this work, we view opinion diversity not as a danger but as a necessity, believing in readers' ability to construct informed opinions.

\subsection{Background: Discord Questions}
\label{sec:background}

We introduce the terminology used in this work to describe news content production, based on the Discord Question framework \cite{laban2022discord}. A \textit{news story} represents an event that happened at a specific point in time (e.g., Sweden and Finland applying to join NATO). The news story receives coverage from \textit{news sources} represented by a domain name (e.g., cnn.com). News sources publish \textit{news articles} that typically cover a single news story. A news article is at a minimum composed of a headline and main content, which can be decomposed into paragraphs. We focus on the news stories that receive coverage from at least 10 distinct news sources, in which exhaustively reading all news articles is prohibitively time-consuming.

In the Discord Question framework, news coverage diversity is defined as the formulation of a question, accompanied by diverse -- and sometimes contradicting -- answers from the sources. The premise of the framework is that each question can serve as an analysis tool to reveal how sources position themselves on a specific aspect of a news story (see Figure~\ref{fig:high_level_examples} for example discord questions). An automatic pipeline is used to generate candidate discord questions and filter them down to a final set of discord questions for a given news story. Each question is paired with a consolidated list of answers from news sources. According to the framework, a question qualifies as a discord question if it qualifies several properties: it must (a) be answered by at least 30\% of the sources of a news story, (b) receive a diverse set of answers (i.e., the majority of answers should not be semantically equivalent), and (c) be specific to the story (i.e., unanswered in other news stories). Property (a) filters overly specific questions (e.g., In what year did the factory first open?), (b) limits consensus factoid questions (e.g., Who is the president of the US?), and (c) filters vague or generic questions (e.g., What did they say?).

Authors of the Discord Questions pipeline experiment with varying models to optimize the pipeline and enable the generation of discord questions for any given news story. The final pipeline is composed of three Transformer-based models: (1) a BART-large\cite{lewis2020bart} a question generation model trained on a combination of the InquisitiveQG  \cite{ko2020inquisitive} and NarrativeQA \cite{kovcisky2018narrativeqa} datasets, (2) a RoBERTa-large \cite{liu2019roberta} extractive question answering model trained on the NewsQA \cite{trischler2017newsqa} and SQuAD 2.0 \cite{rajpurkar2018know} datasets, and (3) a RoBERTa-large  answer consolidation model trained on the MOCHA dataset \cite{chen2020mocha}.

Evaluation of the final pipeline finds that although the automatic generation lags human ability at generating discord questions, the pipeline is able to produce multiple discord questions for any news story containing at least 10 news articles. Manual analysis of the produced questions reveals that discord questions can surface four main types of coverage diversity: (1) differences in the level of detail, (2) differences in the aspect discussed (e.g. a political vs. economics perspective on a question), (3) differences in sentiment, and (4) differences in reasoning.

However, \citet{laban2022discord} did not integrate the questions into news-reading interfaces or evaluate their usefulness to accompanying readers while they read the news. In this work, we use the existing implementation of the pipeline and focus on evaluating its integration into practical news-reading interfaces.

\section{Assembly Interfaces}
\label{sec:system_description}

We designed three novel interfaces that incorporate discord questions in varying ways. We attempted to select designs of different novelty and complexity to explore the space of possible designs: the Annotated Article introduces minimal changes to existing news articles and could be implemented as an add-on to a news website, the Recomposed Article remains similar to a news article in appearance but is generated from scratch, and the Question Grid proposes a new information-dense layout. We also reproduce two baseline interfaces -- the Headline List and the News Article -- that do not present discord questions but represent existing news reading interfaces.

\subsection{Annotated Article}
\label{sec:pres_annotated_article}

\begin{figure*}
    \centering
    \fbox{
    \includegraphics[width=0.7\textwidth]
    {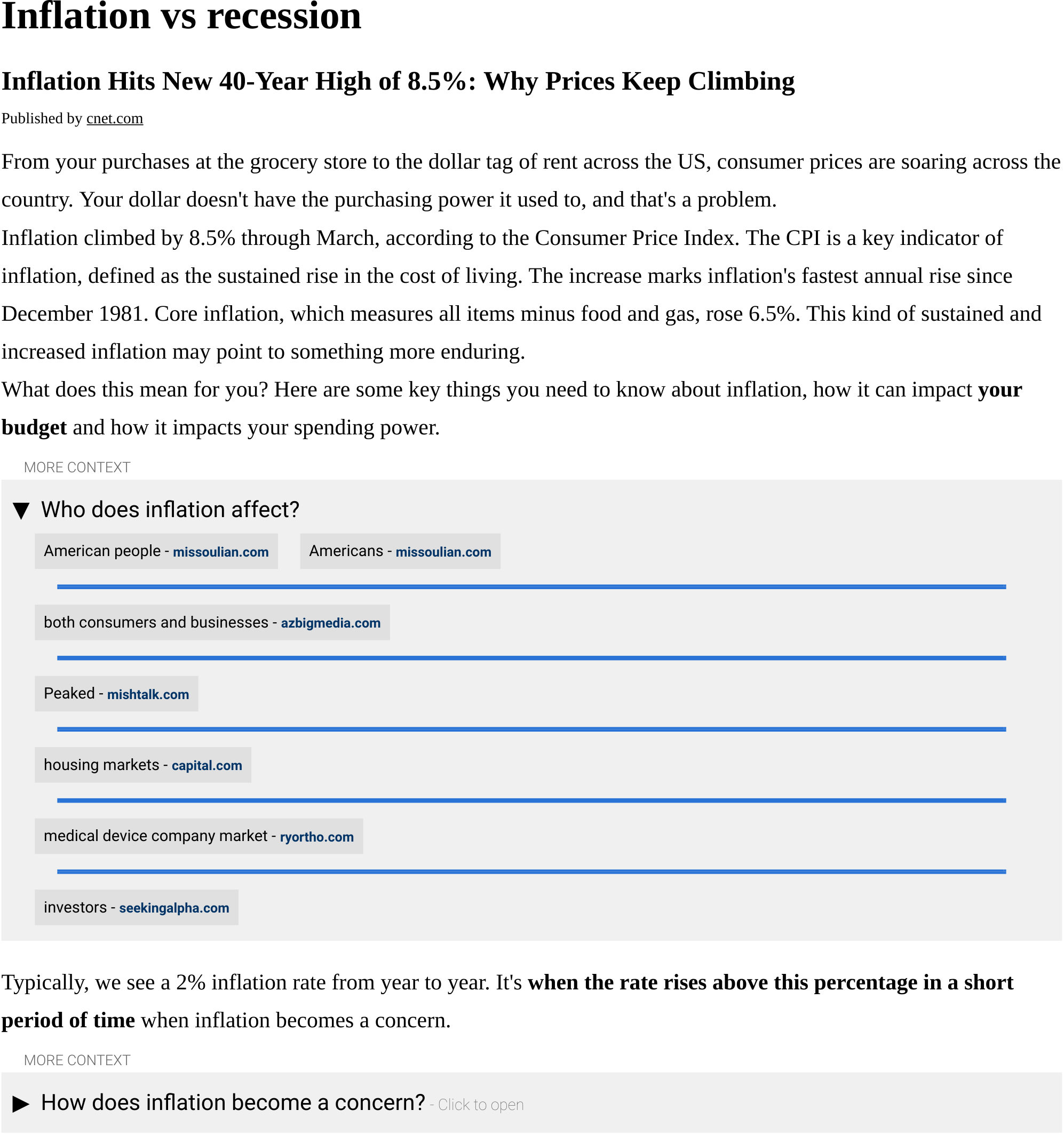}
    }
    \caption{\textbf{Annotated Article interface}. Two discord question-based annotations are inserted into a CNet news article. The first annotation is expanded and the second is collapsed.}
    \label{fig:annotated_article}
\end{figure*}

In the Annotated Article, illustrated in Figure~\ref{fig:annotated_article}, a \textit{basis article} is selected and its contents are reproduced unaltered. Annotations based on discord questions are inserted between paragraphs of the basis article. 

\subsubsection{Design Rationale}

By leveraging a high-quality article as a starting point and adding further annotation, the interface is likely to follow a coherent reading order, and introduce discord questions when they become relevant to the story. The annotations act as an add-on to the basis article, each representing optional additional content, which a user can opt into reading.

\subsubsection{Implementation}

The headline of the basis article is at the top of the interface, followed by a by-line detailing the basis article source, and the sequence of paragraphs of the article. If a paragraph contains the answer to a discord question (i.e., it belongs to the answer group of a discord question), an \textit{annotation} is appended directly after the paragraph. In the example of Figure~\ref{fig:annotated_article}, the second paragraph mentioned that the readers' budget could be impacted by inflation, and the discord question ``Who does inflation affect?'' is inserted as an annotation.

Annotations are rendered as a collapsible rectangular box, togglable through a user's click. When collapsed only the discord question is visible, and once opened a list of all answers to the question from other sources' becomes visible. Each answer is inserted on a separate line, with a clickable link to the original source that supplied the answer. All annotations are collapsed initially. In Figure~\ref{fig:annotated_article}, the first annotation is expanded, and the second collapsed.

For a given news story, we automatically select the basis article by counting the number of annotations each news article would have, picking the article with the most annotations. In the extreme, each paragraph of the basis article is associated with a discord question, and the Annotated Article alternates between paragraphs and annotations.

\subsubsection{Expected Benefits}

\textbf{Familiarity}. An expected benefit of the interface is the ease of use, as a user can choose to not expand the annotations, reducing the interface to reading the basis article, an interface most news readers should be familiar with.

\textbf{Coherence of Text Order.} Because the interface adopts the order of the human-written basis article, which likely follows journalistic guidelines, introducing the required background as needed for an average reader. Because the annotations are placed following the content that they are most related to within the article, they are likely to minimally affect content coherence.

\subsubsection{Expected Drawbacks}

\textbf{Limited Coverage}. The Annotated Article is limited in the number of discord questions presented to the user, as only questions addressed within the basis article are appended as annotations. For example, the framework might produce a total of 20 discord questions, but the best basis article might only accommodate eight questions, and the user will not be exposed to an additional 12 discord questions.

\textbf{Low Diversity by Default}. Because all annotations are collapsed by default, a second drawback is that without additional user effort, the majority of content originates from a single source -- the basis article -- reducing coverage diversity. Although it is possible to expand the annotations by default, initial feedback from users showed that could be too disruptive, and collapsing the annotations by default gives the user more control on alternating between reading the basis article and the annotations.

\subsection{Recomposed Article}
\label{sec:pres_recomposed_article}

\begin{figure*}
    \centering
    \fbox{
    \includegraphics[width=0.65\textwidth]{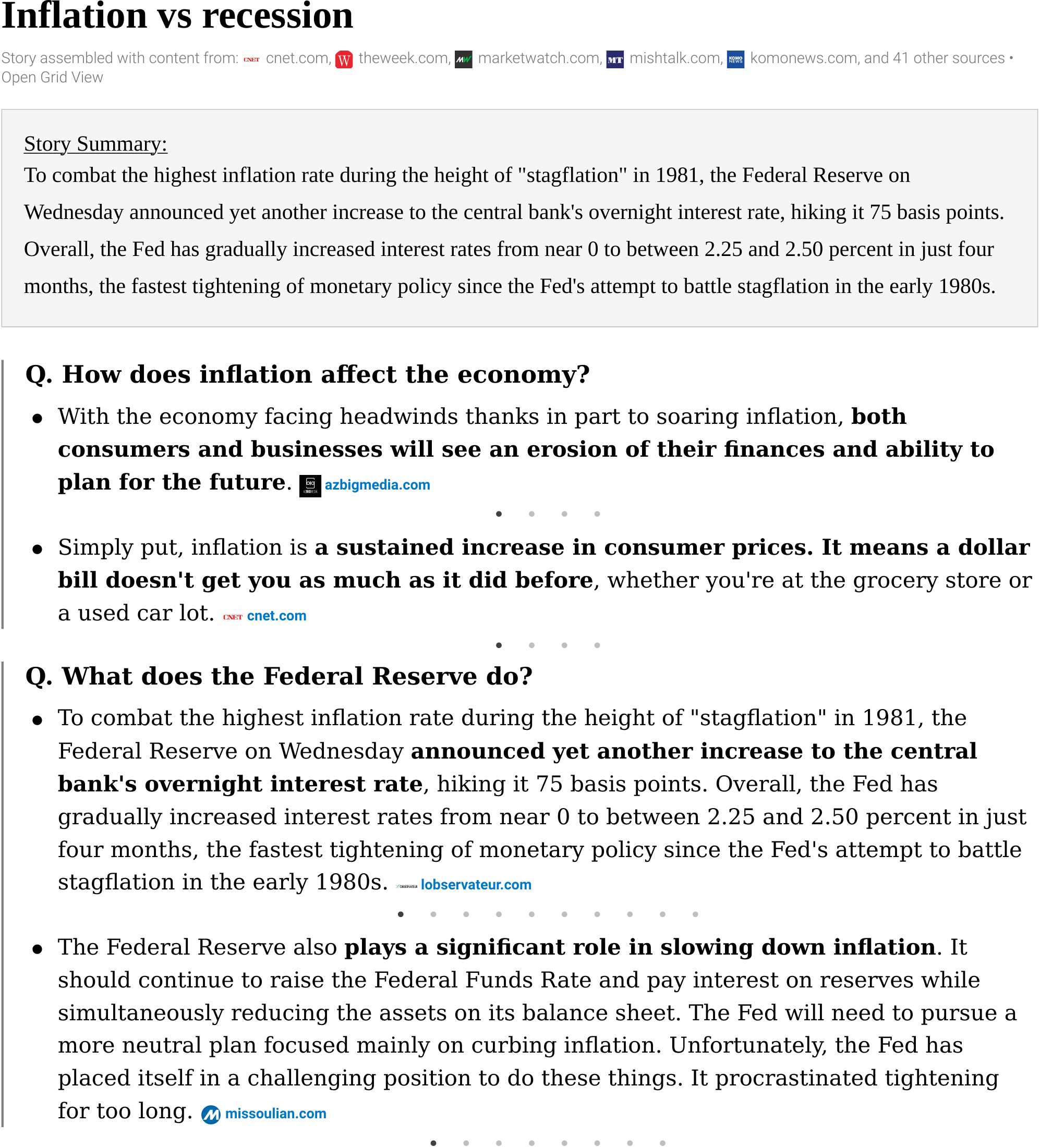}
    }
    \caption{\textbf{Recomposed Article interface}. In the upper portion, the sources used are listed, followed by an introductory summary, and a sequence of discord questions with corresponding answer paragraphs.}
    \label{fig:recomposed_article}
\end{figure*}

In the Recomposed Article, illustrated in Figure~\ref{fig:recomposed_article}, an article is created de novo using the content of several news sources. First, a summary is extracted from one of the source articles, intended to present the basic facts and context of the news story. The second portion consists of a sequence of discord questions, each composed of the question itself followed by a list of paragraphs containing answers to the question from the story's sources. 

\subsubsection{Design Rationale}

The Recomposed Article keeps the overall layout of a textual document meant to be read from top to bottom to resemble a standard news article. The two-step reading process, first introducing necessary minimal context followed by in-depth content, follows the inverted pyramid style common in journalism \cite{po2003news}.

\subsubsection{Implementation}

The upper portion of the interface introduces the story name, followed by a by-line of all source articles used in the composition.

\textbf{Summary Selection}: News articles often include a manually written summary within the page's metadata \cite{grusky2018newsroom}. We extract summaries from all sources and select one closest in length to 60 words. In practice, we find that this simple heuristic yields summaries that give an appropriate introduction to the story.

\textbf{Sequence of Discord Questions}: A composition algorithm (pseudo-code in Appendix~\ref{fig:algo_recomposed_article}) is used to select and order discord questions. At a high level, the algorithm iterates over discord questions, selecting questions that introduce the largest amount of unseen content. With this algorithm, questions that are addressed by more sources tend to come earlier in the sequence, and more specific questions discussed by fewer sources appear later. This ordering mirrors the inverted pyramid writing style.

Visually, each selected discord question is represented by a rectangular unit in the interface. The question is followed by two-paragraph answers from distinct sources, in bullet-point format. For each paragraph, the answer span is bolded. In cases where more answers are available, they are added as a horizontal carousel, allowing the user to dig deeper when interested. Source attribution is appended to each paragraph as a blue link, allowing the user to easily access the source of an answer of interest.

\subsubsection{Expected Benefits}

\textbf{Article Appearance.} The Recomposed Article maintains the appearance of a news article, with a top-to-bottom reading direction, and an inverted pyramid writing style, which should be familiar to news readers.

\textbf{Full Paragraphs.} The Recomposed Article is the only Assembly interface to present full-paragraph answers to discord questions, providing additional context in cases when spans alone can be hard to interpret.

\subsubsection{Expected Drawbacks}

\textbf{Lack of Coherence.} Since the order of Discord Questions is chosen algorithmically, the overall article will likely lack thematic coherence. \textbf{Excessive Length.} The length of the Recomposed Article could be excessive, particularly for stories with a large number of discord questions. The unabridged version of the interface shown in Figure~\ref{fig:recomposed_article} contains 39 discord question units and approximately 3,900 words.

\subsection{Question Grid}
\label{sec:question_grid}
\begin{figure*}
    \centering
    \fbox{
    \includegraphics[width=0.9\textwidth]{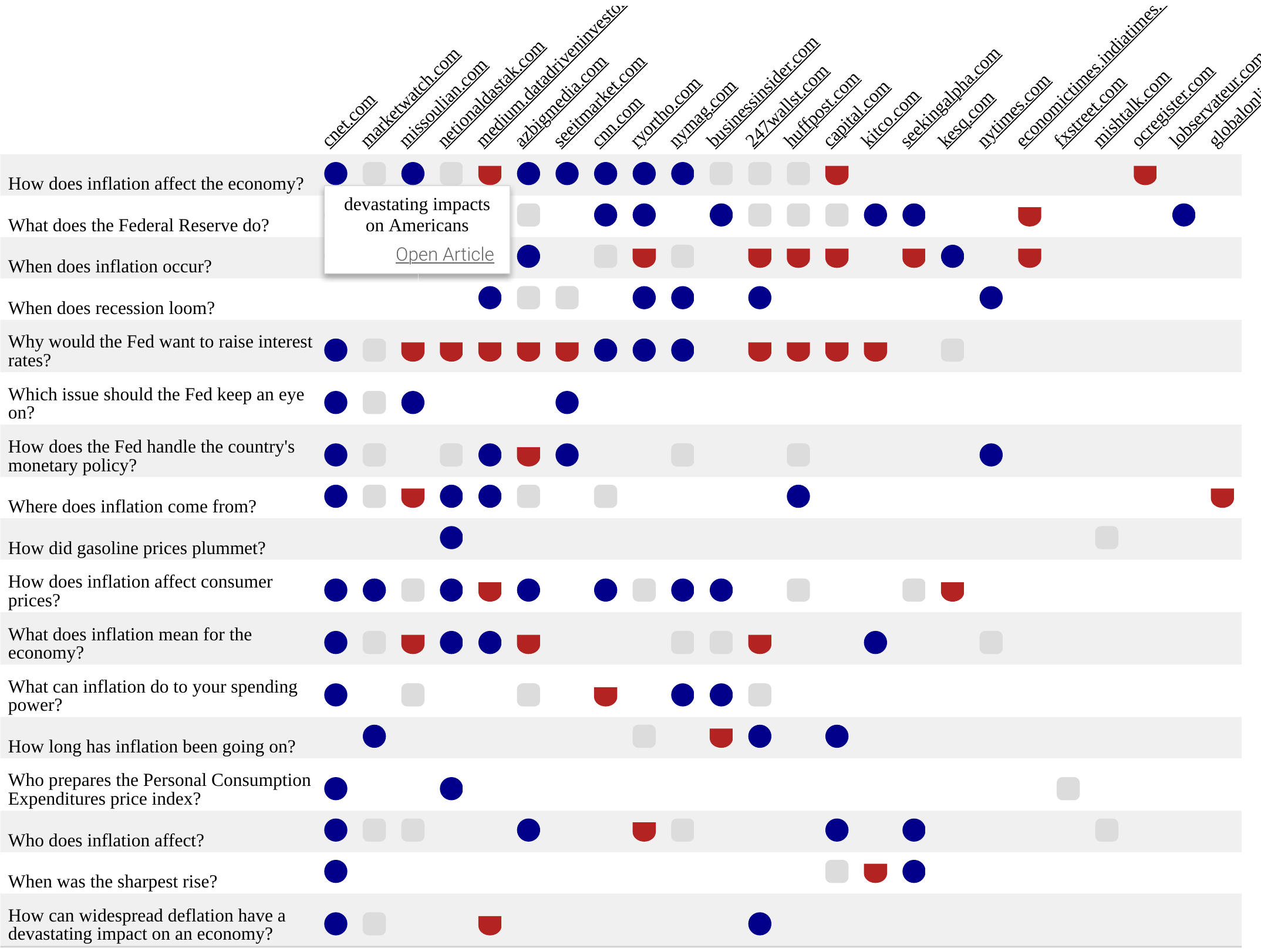}
    }
    \caption{Question Grid interface for a news story on ``Inflation / Recession in 2022''. Each row lists a question, each column represents a source. For each element in the grid, the presence of a square indicates a source answering the row's question, and the color indicates answer similarity. Hovering over a square displays the source's answer.}\vspace*{-6pt}
    \label{fig:question_grid}
\end{figure*}

In the Question Grid, illustrated in Figure~\ref{fig:question_grid}, the news story is rendered as a two-dimensional grid. Each row of the grid represents a question and each column a source. Each \texttt{(i,j)}-element in the grid is either empty if source \texttt{j} did not answer question \texttt{i}, or a colored shape when an answer is found.

\subsubsection{Design Rationale}

Inspired by prior work \cite{Hamborg2017MatrixBasedNA} leveraging grid-based visualization of news stories, we adapt the discord questions data to the grid format. The information-dense visualization is intended for advanced users, to help compare and contrast sources, and inspect the framing choices of newsrooms.

\subsubsection{Implementation}

The grid representation relies on choosing an order for the questions and the sources. Question ordering -- similarly to the Recomposed Article -- is based on the number of source answers to each question, such that most answered questions are in the upper portion of the grid. Sources are ordered based on the number of questions they answer, with the sources that answer more questions in the left portion of the grid. The combined orders result in the upper-left corner of the grid being the most populated, and other areas of the grid gradually losing answer density.

When source \texttt{j} answers question \texttt{i}, a colored shape is inserted in element (i,j) of the matrix. A hover window appears when the user moves the mouse over answer shapes, containing the answer span of the source \texttt{j} for question \texttt{i}. The user can click on the hover window to open the source's article in a new browser tab.

Color and shape indicate semantic similarity between answers. The Discord Questions framework organizes answers to a question into groups, such that all answers within a group relay similar answers to the question. In the grid, we assign each answer group to a distinct color and shape. For example, the first row of the Question Grid in Figure~\ref{fig:question_grid} corresponds to the question: ``How does inflation affect the economy?''. The seven blue shapes relay that inflation leads to negative effects on consumers, the three red shapes that it leads to an overheating of the economy, and the five grey that it affects the housing market.

In order to reduce redundancy in the grid, questions that overlap in their answering groups (i.e., the paragraphs that answer the question) by more than 80\% are deduplicated, keeping only the question with the larger number of answers. Deduplication ensures that each question brings on a unique axis of diversity within the story. We note that it is likely that the composition algorithm of the Recomposed Article and the deduplication process of the Question Grid select similar questions, leading to the underlying content of the two interfaces to be alike, and the differences exposed in the Usability Studies arising from the presentation of the content.

\subsubsection{Expected Benefits}

\textbf{Information Density.} The Question Grid is the most efficient interface of the three Assembly interfaces in terms of information density. This can be beneficial for users looking to scan for questions of interest, or looking to analyze the coverage of a particular source, as the rectangular layout enables horizontal and vertical inspection. \textbf{Source Comparisons.} The matrix-based interface lends itself to the pairwise comparison of sources. For example using Figure~\ref{fig:question_grid}, it can be visually deduced that BusinessInsider and CNet provide similar coverage of the story, as they answer eleven common questions, whereas MarketWatch and BusinessInsider are more dissimilar, with only five questions in common. \textbf{Answer Omission.} The Question Grid is the only interface to explicitly visualize a source not answering a question, represented by a blank entry in the matrix. As pointed out by prior work\cite{cooper1994sins,ehrhardt2021omission}, surfacing omissions in coverage is an important signal in analyzing source bias.

\subsubsection{Expected Drawbacks}

\textbf{Lack of Context.} The Question Grid interface does not provide a high-level introduction to the story, potentially causing confusion in users unfamiliar with the story. \textbf{Information Overload.} The benefits of information density come at a cost, and it is likely that a densely packed Question Grid might be challenging to some users, due to the overwhelming choice of questions and sources. The requirement to hover over individual answer elements could prove tedious to some users as well.

\subsection{Baseline Interfaces}

In order to expand the field of comparison during our usability studies, we implement two baseline interfaces: the News Article and the Headline List.

\subsubsection{News Article.} The News Article corresponds to the unannotated content of a single news article. This interface is equivalent to the Annotated Article with zero annotations. This basic interface is intended to simulate a user reading content from a single source, offering a baseline of coverage diversity from a single source. We take the article of median length amongst available sources, with an objective to represent the average level of coverage of a single news article (rather than the extrema of longest and shortest articles). We do not include a screenshot of the News Article, as it simply corresponds to the Annotated Article in Figure~\ref{fig:annotated_article} without any annotations boxes.

\subsubsection{Headline List.} The Headline List iterates over each source presenting solely the headline of the article. A user can then click on the headline to open the full news article associated with the source and headline. This interface is common to news aggregators such as Google News or Yahoo News, with prior studies showing that headline-based interfaces reduce the fraction of participants that delve deeper into news stories beyond the headline \cite{rosenstiel_media_2014}. Headline List is intended to simulate users of standard news aggregators. Appendix~\ref{appendix:baseline_interfaces} provides a screenshot of the Headline List.

\section{System Implementation}
\label{sec:system}

We introduce the data source and computational resources to build the live version of the Assembly interfaces.

\subsection{Data Source}

When exposing readers to diverse opinions, we have a responsibility to limit the visibility of harmful sources that introduce misrepresentations of important events, which can lead to manipulation of public opinion\cite{kuypers2006bush,allcott2017social}.

There is an engineering challenge in maintaining a list of trusted news sources. Google News -- the most popular news aggregator in the US according to Pew Research\cite{pewresearch2020} -- bases its news recommendation on content from more than 50,000 sources \cite{filloux_2013}.

Google recently released a document describing the principles behind the source selection for Google News\cite{googledev2021}, outlining the manual review process and the editorial expectations for sources within its collection.

In our prototypes, we rely on Google's source selection process, acquiring groups of diverse sources covering a common story directly from the live Google News website. We programmatically visit the Google News pages for World, Finance, Politics, Business, and Science sections, extracting each news story with at least 10 distinct news sources.

For each story, we then directly access each of the source articles and use the \texttt{newspaper}\footnote{\href{https://github.com/codelucas/newspaper}{github.com/codelucas/newspaper}} library to extract the plain text article. In some cases, an article can only be accessed with a paid account, in which case we only extract basic metadata such as the headline and summary when available.

Although we depend on Google's source selection process, it is not a gold standard and is known to have Western bias \cite{Watanabe2013TheWP}, for instance with the aggregator recently removing major Russian sources from its platform\footnote{\href{https://www.reuters.com/technology/exclusive-google-drops-rt-other-russian-state-media-its-news-features-2022-03-01/}{reuters.com/technology/google-drops-rt-other-russian}}. Striking the right balance between increasing access to opinion diversity and minimizing harmful content is fundamentally challenging.

\subsection{Computational Resources}

The Discord Question pipeline processes Google News stories as they are published. On average, the stories contain 37 news articles. QGen takes as input individual news articles and generates questions, producing on average around 987 candidate discord questions per story. QA takes as input each candidate's discord question paired with each news article and extracts an answer when one is found. A third and final process confirms or discards each candidate question, based on whether it receives answers from enough sources and whether the answer set is diverse. On average, the pipeline produces 16 discord questions.

We run the Discord Questions pipeline on a single server equipped with 4 Nvidia V100 GPUs, one allocated to QGen, two to QA, and one to filtering. With the described resources, we are able to process the incoming stream of stories from Google News, on average processing 403 stories per day.

\section{Usability Study A: Journalism Experts}
\label{sec:experts_study}

We conducted a usability study with 10 journalism professionals. The objective were to:
\begin{enumerate}
    \item Understand potential use-cases of each Assembly interface,
    \item Determine how the Assembly interfaces compare to baseline interfaces in terms of ease of use and coverage diversity presentation,
    \item Assess which of the interfaces are suitable for use by experts and/or novice news readers.
\end{enumerate}

\subsection{Participants}

We recruited 10 participants (5 women, 4 men, and 1 non-binary, aged between 23 and 71, all living in the US) on a user research recruiting platform\footnote{\url{https://www.userinterviews.com}}. Participants were recruited based on a screener survey, with an objective to recruit journalism professionals with diverse experience. In the screener survey, participants reported having 1-35 years of experience in journalism (mean 10.4 years), and working at both local news organizations (e.g., Arizona Pbs, Laredo Morning Times) and national news organizations (e.g., CNN, NBC News). In terms of roles within the newsroom, 9 participants listed reporting/writing, 4 listed editing, and 2 producing. Finally, participants listed different specializations, from Local Politics to Crime \& Courts, and Technology.

\subsection{Study Procedure}

The study was conducted in 1-hour sessions online via video-conferencing software, with participants receiving \$80 upon completion. The interface and study procedure were approved by the ethics review agent of our organization.

Each participant first watched a 3-minute introductory video introducing terminology and features of the five interfaces (2 baselines, 3 Assembly). Participants then completed two 25-minute story-reading sessions each involving all interfaces, and finally completed an open-ended feedback form.

The two story-reading sessions were centered on major US news stories at the time of the study: the Baby Formula Shortage, and Potential Recession. Each story-reading session consisted of interface browsing and a comparison questionnaire. During interface browsing, participants spent 4 minutes viewing the story with each interface, going in the following order: News Article, Headline List, Annotated Article, Recomposed Article, Question Grid. Participants were instructed to browse the interface with the objective to learn about the story and were permitted to click on external links when available.

\begin{figure*}
    \centering
    \begin{subfigure}[b]{0.39\textwidth}   
        \centering 
        \includegraphics[width=\textwidth]{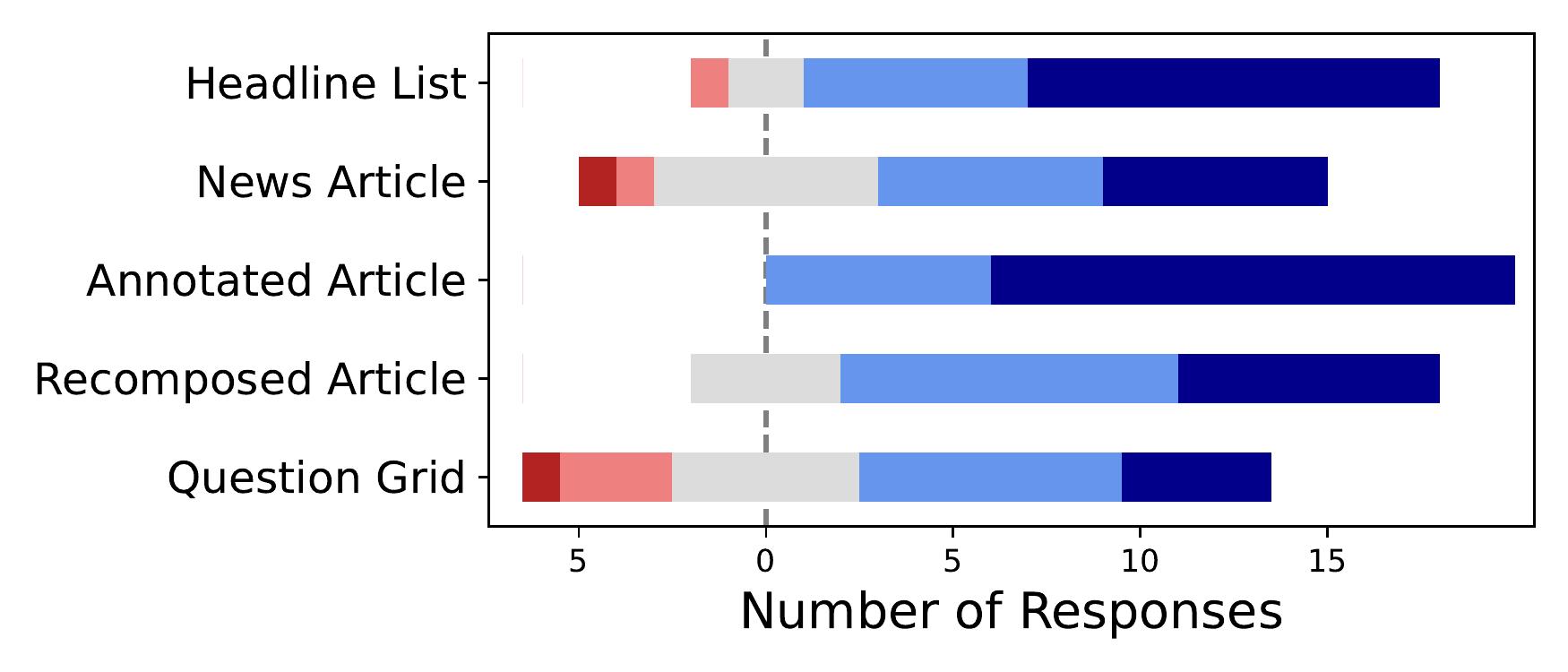}
        \caption[]{Gives \textbf{good overview} of story}
        \label{fig:exp_good_overview}
    \end{subfigure}
    \hfill
    \begin{subfigure}[b]{0.30\textwidth}   
        \centering 
        \includegraphics[width=0.92\textwidth]{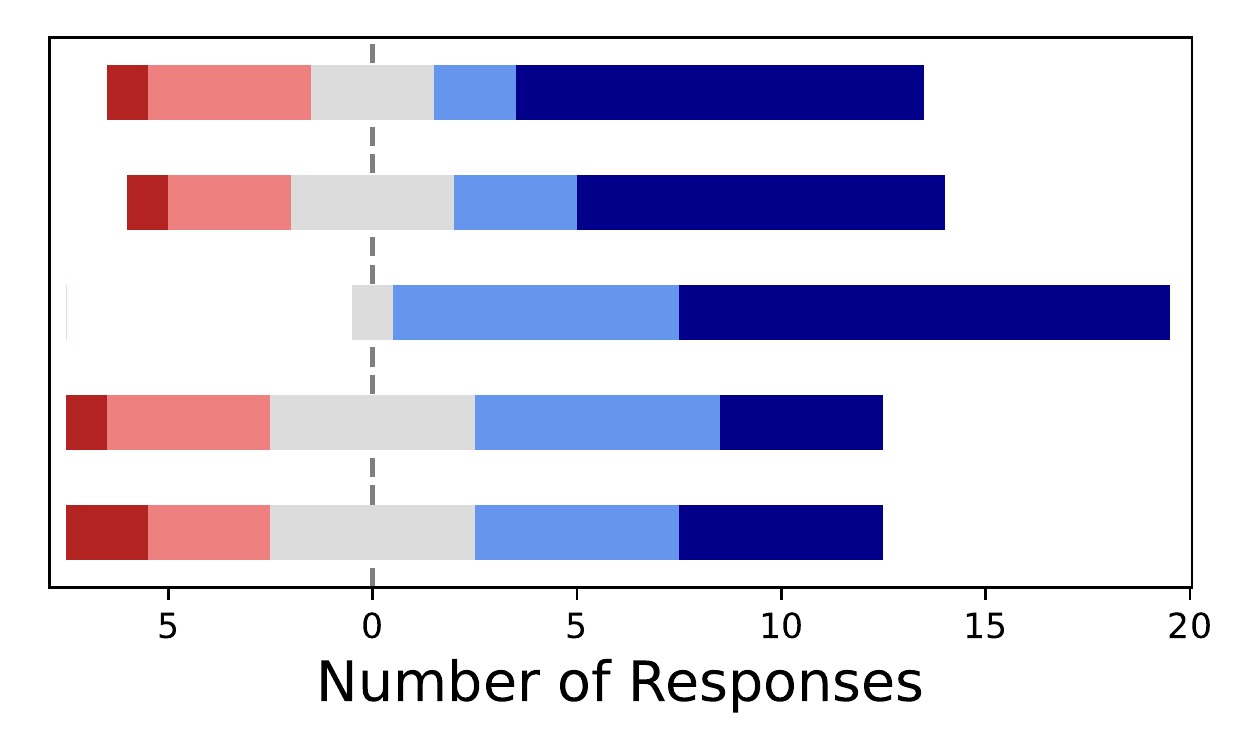}
        \caption[]{Information is in \textbf{coherent order}}
        \label{fig:exp_coherent}
    \end{subfigure}
    \hfill
    \begin{subfigure}[b]{0.30\textwidth}   
        \centering 
        \includegraphics[width=0.92\textwidth]{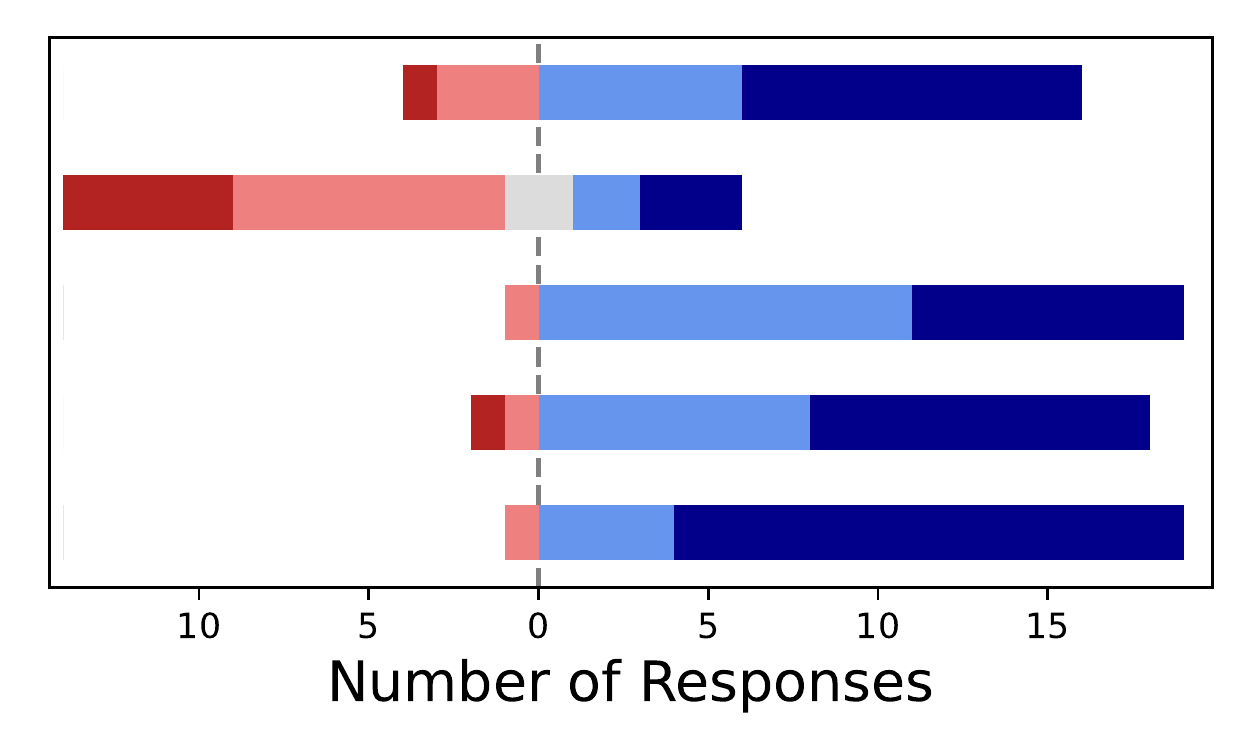}
        \caption[]{Reveals \textbf{coverage diversity}}
        \label{fig:exp_coverage}
    \end{subfigure}
    \hfill
    \vskip\baselineskip
    \hfill
    \begin{subfigure}[b]{0.384\textwidth}
        \includegraphics[width=\textwidth]{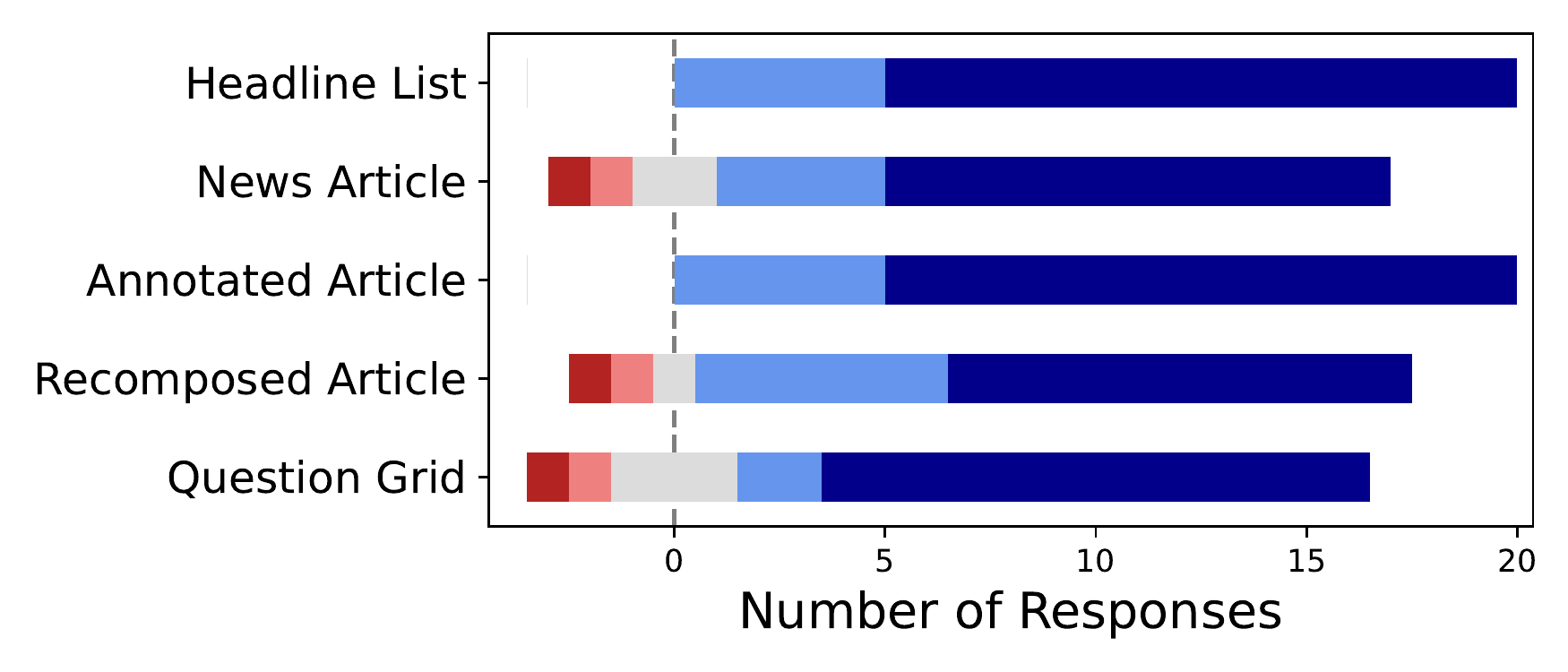}
        \caption[]{Is easy to use for \textbf{media experts}}
        \label{fig:exp_easy_expert}
    \end{subfigure}
    \begin{subfigure}[b]{0.50\textwidth}
        \includegraphics[width=0.81\textwidth]{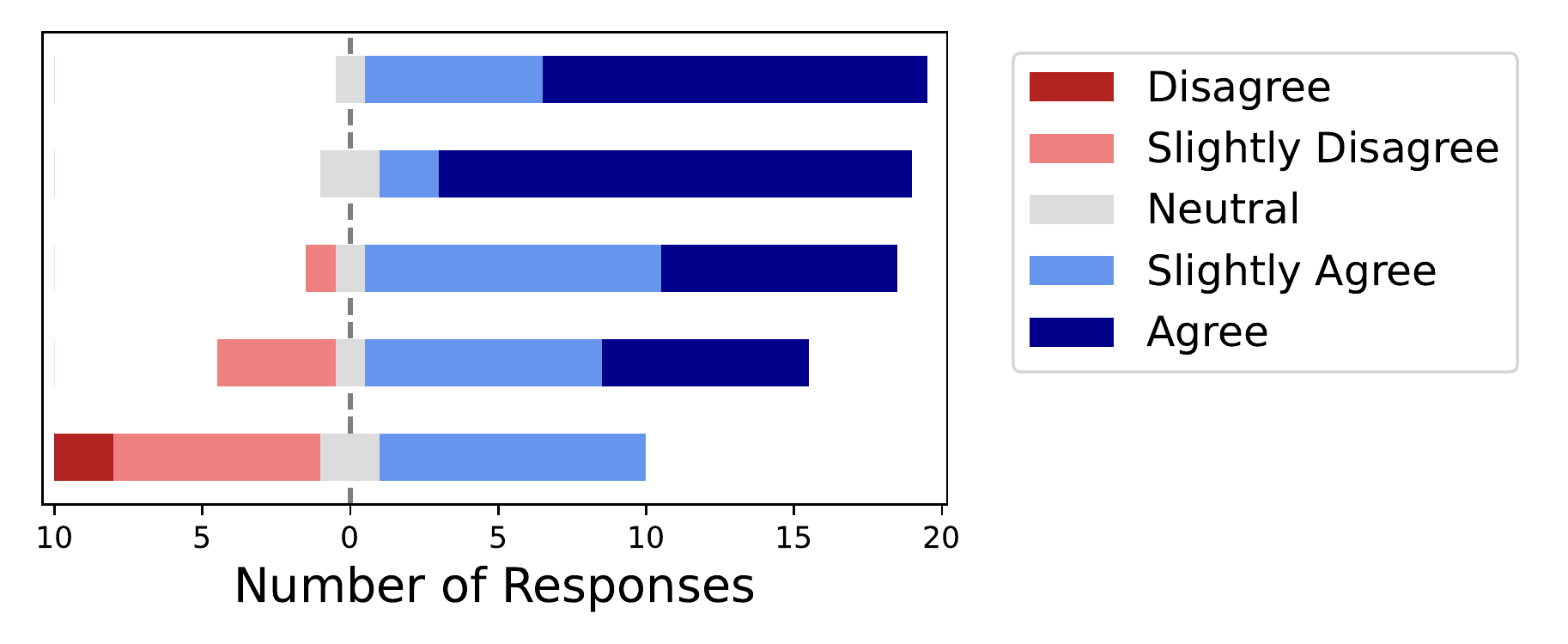}
        \caption[]{Is easy to use for \textbf{novice readers}}
        \label{fig:exp_easy_novice}
    \end{subfigure}
    \hfill
\caption[]{\textbf{Expert Comparison Questionnaires Results.} Participants rated  each interface on whether it: (\hyperref[fig:exp_good_overview]{a}) provides a good overview of the story, (\hyperref[fig:exp_coherent]{b}) presents information in a coherent order, (\hyperref[fig:exp_coverage]{c}) reveals coverage diversity, (\hyperref[fig:exp_easy_expert]{d}) is easy to use by experts, and (\hyperref[fig:exp_easy_novice]{e}) easy to use by novice users.}
    \label{fig:expert_likerts}
\end{figure*}

In the comparison questionnaire, participants provide 5-pt Likert scale ratings of the interfaces. Rather than requesting ratings after using each interface, we waited until participants had seen all interfaces and used a grid of Likert scales to allow participants to rate interfaces relative to each other. Before they provided ratings, we confirmed with each participant that they were familiar with each interface's name, and allowed participants to go back to individual interfaces as they answered. Figure~\ref{fig:expert_likerts} summarizes comparison questionnaire results.

\subsection{Study Results}

\subsubsection{Revealing Coverage Diversity (Figure~\ref{fig:exp_coverage})} A larger proportion of experts found that the Assembly interfaces revealed coverage diversity in the story than the baseline interfaces. The Question Grid achieved the highest ratings, followed by the Recomposed Article and the Annotated Article. The single News Article was rated largely lower, confirming that content from a single source does not provide satisfactory coverage diversity for complex stories. The Headline List -- a popular interface in existing news aggregators -- obtains ratings between the News Article and the Assembly interfaces, confirming that it is partly successful at conveying coverage diversity.

The Discord Questions-based annotations are the only difference between the News Article baseline and the Annotated Article, yet lead to a large rise in ratings. This difference highlights that a news article can provide support to introduce a news story, and annotations can be inserted to increase coverage diversity.

The Recomposed Article and Question Grid achieve the highest ratings, with the latter slightly better rated. Although both interfaces relay largely the same information (see Section~\ref{sec:question_grid}), we hypothesize that the increased information density of the grid gave experts a perception of better access to coverage diversity.

\subsubsection{Good Overview \& Coherent Order (Figure~\ref{fig:expert_likerts}\hyperref[fig:exp_good_overview]{a}-\hyperref[fig:exp_coherent]{b})} The Annotated Article achieves the highest ratings in terms of both coherence and providing an accessible overview of the story. This confirms that the method of adding annotations directly after the related paragraph in a human-written news article is the least disruptive way to introduce discord questions content without reducing story accessibility.

The two more advanced Assembly interfaces -- Recomposed Article and Question Grid -- obtain worse ratings overall, with the Question Grid lagging other interfaces in terms of giving a good overview of the story. These ratings validate that the pair achieves higher coverage diversity at the cost of accessibility.

The two baseline interfaces are rated between the advanced Assembly interfaces and the Annotated Article, and are overall seen as providing good introductory material to news stories.

\subsubsection{Ease of Use (Figure~\ref{fig:expert_likerts}\hyperref[fig:exp_easy_expert]{d}-\hyperref[fig:exp_easy_novice]{e})} Expert participants were asked to estimate whether the interface would be easy to operate for other media professionals, as well as for novice news readers without expertise in journalism.

With respect to an expert population, all interfaces received high ratings, with 80\% of experts agreeing that fellow experts should be able to use any of the interfaces. The most information-dense interface -- the Question Grid -- received slightly lower ratings, coming from two participants that found the grid hard to understand (more detail in qualitative feedback \S\ref{sec:exp_qualitative}).

With respect to a novice reader population, rating distributions showed a larger variance. Only three interfaces were considered suitable by a large majority of participants: the two baselines and the Annotated Article. The Recomposed Article and Question Grid both received lower ratings, with no participant strongly agreeing that the Question Grid is suited for novice users. We hypothesize that lower ratings are due to the two interfaces not being anchored on a news article and as such not providing the introductory context that novice readers might require.

\subsection{Qualitative Feedback}
\label{sec:exp_qualitative}

Participants were given time to provide open-ended feedback and were asked to reflect on: (1) the limitations or benefits of any interface, (2) potential use cases of any interface, and (3) any aspect that did or did not work well. We employed a thematic analysis \cite{braun2012thematic} to organize the feedback, and we discuss themes brought up by three of the ten experts or more.

\subsubsection{Preference for the Annotated Article -- 8 participants} Eight of the ten participants explicitly expressed an overall preference for the Annotated Article, mirroring the interface's high ratings in Figure~\ref{fig:expert_likerts}.

\begin{quote}
    \textit{``I really thought the annotated article was well done.''} -- (P10)
\end{quote}

We observed during the interviews that participants interacted differently with the annotations, with some choosing to first read the article, followed by opening some annotations, and others choosing to open each annotation as they appeared in the article. Overall, participants often expanded 50-80\% of the annotations of the article to see multi-source answers.

\subsubsection{News Article does not provide coverage diversity -- 5 participants} Half the participants noted the lack of coverage diversity in the single news article baseline interface. This finding serves as a justification for the other four interfaces which integrate multi-source content to achieve higher coverage diversity.

\begin{quote}
    \textit{``I did not see a lot of use for the single article -- I am not interested in having my reading tailored that specifically.''} -- (P5)
\end{quote}

\subsubsection{Question Grid Use Cases -- 5 participants} First, several participants saw an opportunity in the blank spaces of the Question Grid. By seeing the parts of the Grid that lack coverage, a newsroom could decide on story angles to produce stories that are differentiated from the competition.

\begin{quote}
    \textit{``I could see journalists using the question grid as a way to inspire different angles in stories, as well as start researchers down different paths of questioning.''} -- (P3)
\end{quote}

Second, some participants proposed processing the grid itself to inspect story coverage. By analyzing which questions are more or less answered, a journalist could computationally discover the frames of a news story.

\begin{quote}
    \textit{``I can also see the question grid being used as a kind of data source to make an argument about news diversity itself''} -- (P1)
\end{quote}

However, two participants expressed difficulty in using the Question Grid, finding the interface sometimes overwhelming, which would require better training to operate.

\begin{quote}
    \textit{``The grid was a bit difficult to understand at first. It's a good tool but needs a bit more of explanation.''} -- (P8)
\end{quote}

\subsubsection{Discord questions can be noisy -- 4 participants} Although participants were not told the questions and answers were automatically generated, some spotted noisy elements in the questions or answers.

\begin{quote}
    \textit{``Some questions that were visible in the interface had inaccurate or imprecise ``answers'' listed''} -- (P9)
\end{quote}

To counter the imperfections, one participant suggested manual editing would be required before publishing.

\subsubsection{Trustworthiness of Sources -- 2 participants}

We further spotlight two insights from individual participants that are relevant to future work and the limitations of our work. First, during the study, P5 was visibly irritated by the opinion of certain sources that were being highlighted in some interfaces. In their feedback, P5 mentions the importance of the user's trust in the sources included in the interface:

\begin{quote}
    \textit{``I need assurance that the search algorithm that produces the stories I see is not slanted toward any particular results and has a reasonably broad scope. [...] Again, that is only useful if I trust the sources.''} -- (P5)
\end{quote}

By leveraging the source selection process of Google News, we ensure some level of quality of included sources. Source trust is however user-specific, and an empowering solution for future work would be to give the user control over source inclusion, allowing them to filter out undesired sources.

On a related note, participant P10 pointed out the danger of plurality of answers at the center of the Discord Questions framework when a single answer is correct, again hinting at trust in sources being important:

\begin{quote}
    \textit{`The downside to that [discord questions] might be all the different answers you get. It could be confusing for a reader to know which source is the trusted one with several different answers.''} -- (P10)
\end{quote}

Diversity in answers is valuable in some cases but can be detrimental when it is used to spread misinformation. Providing the user information about included sources and control to remove unwanted sources could be implemented in future work, reducing reliance on Google's source selection process.

\subsection{Summary of Results}

In summary, only three of the five interfaces were estimated to be operable by novice news readers. The only Assembly interface to make the cut -- the Annotated Article -- was particularly well received, as it was rated to provide the best overview of the story while highlighting almost as much coverage diversity as more advanced interfaces.

The two advanced Assembly interfaces -- Recomposed Article and Question Grid -- maximize exposure to coverage diversity at the cost of ease of use and are limited to subject-matter experts looking to deepen their understanding of a story. Surprisingly, although the Question Grid was rated as slightly more challenging to use, participants were more vocal about its potential use cases in their open-ended feedback. The Recomposed Article -- which can be thought of as a middle-ground between the other two Assembly interfaces -- did not find its preferred audience, as the Annotated Article was a favorite for introductory use cases and the Question Grid for advanced analysis.

Following the recommendation of the participating experts, we exclude the Recomposed Article and the Question Grid interfaces from the usability study involving novice news readers, focusing on the three interfaces that were predominantly rated as easy to use by novice news readers.

\section{Usability Study B: Novice News Readers}
\label{sec:novice_study}

In a second usability study with novice news readers, the objectives were to:
\begin{enumerate}
    \item Assess whether the Annotated Article leads readers to gain a broader understanding of a news story compared to baseline interfaces,
    \item Verify whether the Annotated Article is as straightforward to use as baseline interfaces,
    \item Understand user pain points in using the News Article, the Headline List, and the Annotated Article.
\end{enumerate}

Measuring reader exposure to coverage diversity is challenging \cite{Spinde2020EnablingNC}, prompting us to design an active reading exercise to testbed interfaces (\S\ref{sec:exercise_design}). We then detail the study protocol (\S\ref{sec:nov_procedure}), report on recruited participants (\S\ref{sec:nov_participants}), detail the manual analysis we conducted (\S\ref{sec:manual_scores}), and analyze quantitative and qualitative outcomes (\S\ref{sec:nov_results}-\ref{sec:nov_qualitative}).

\subsection{Reading Exercise Design}
\label{sec:exercise_design}

We pose three requirements for the study design. Requirement I: the study should simulate a realistic reading scenario for a novice reader (i.e., it should not require participants to use an interface for several hours, or require advanced training). Requirement II: the study should strive to be topically interesting to the user to maximize genuine interaction from the user while being relatively novel to minimize the effects of prior knowledge from participants. Requirement III: the study should be reproducible and not provide an unfair advantage to one of the settings. We introduce the proposed study design, explaining choices made with respect to the target requirements.

\textbf{Overall exercise.} The exercise is a time-limit reading comprehension exercise consisting of four open-ended questions. During the exercise, a participant has six minutes to answer the questions using a single interface, assigned at random. The participant is urged to answer the questions in the bullet-point form and as thoroughly as possible, listing any answer elements they read.

\textbf{Concurrent reading and answering.} A major design choice is whether the participant completes the comprehension questions during or after the reading session. We choose a concurrent design using a two-column interface shown in Appendix~\ref{appendix:exercise_interface}. The advantage of the concurrent design is that it does not rely on participants recalling answers, and allows participants to actively use the interface with an objective, avoiding passive navigation (requirement II).

\textbf{Story Selection.} Participants are given story choices and are prompted to select a story they are interested in but not up to date on. It is important that the participant is both interested to generate genuine interaction (requirement II), without having participants with too much prior knowledge that would bias results (requirement III). 

\textbf{Unbiased Question Selection.} Questions selection should be interface-independent. For example, the questions should not be taken from the discord questions in the Annotated Article, as this would bias results and reduce reproducibility (requirement I). As further detailed in Section~\ref{sec:nov_procedure}, we hired external experts to write questions based on their review of the study's selected news stories.

\textbf{Time-limited Exercise.} The average reading time for long-form news articles in 2016 was a little over 2 minutes \cite{pewresearch2016}. By setting the exercise duration at 6 minutes, we account for the additional time needed to write down answers and give participants roughly the reading time of a realistic news-reading session (requirement I). Once the time has elapsed, a message is shown and participants are given up to three additional minutes to finalize their answers.

\textbf{Open-ended Short-form Questions.} We select comprehension questions that require reasoning, prediction, and generally short-form answers, rather than factoid questions that would focus the exercise on narrow fact-finding. We did not impose within the interface the use of bullet points, as we performed manual scoring of the answers (see more in \S\ref{sec:nov_procedure}) and can in this way detect low-quality participants.

\textbf{Allowing no-ans.} Participants were explicitly told it is acceptable to leave an answer blank or answer it with ``No answer''. Although we expected most questions to be \nobreak answerable with the three interfaces, we believe that participant perception of a lack of \nobreak answers is an interesting signal. Some participants might abuse the ability to leave answers empty, and we discuss how we filtered out such participants in \S\ref{sec:nov_participants}.\enlargethispage{16pt}

\subsection{Study Procedure}
\label{sec:nov_procedure}

\subsubsection{Study Content}

We selected the two expert study stories -- Baby Formula Shortage, and Potential Recession -- and three new stories: New Iran Nuclear Deal, Taiwan/China Diplomatic Row, and Sweden/Finland Joining NATO.

To generate comprehension questions, we re-contacted three experts from the first usability study. Each expert spent at least 10 minutes getting informed on each story, and was tasked with writing 4-6 questions (without answers). Question-writers were asked to favor core questions with multiple answer aspects or opinions. Through this process, we obtained roughly 12 questions per story and selected a final four by deduplicating and prioritizing questions suggested by several experts. As an example, comprehension questions for Taiwan/China Diplomatic Row are: (1) ``How is China responding to separatists in China?'', (2) ``How is Taiwan responding to Chinese threats?'', (3) ``Why was Nancy Pelosi visiting Taiwan?'', (4) ``What did the Chinese government say in a white paper?''.

We manually verified that no selected question matched verbatim to any of the discord questions within the Annotated Article, confirming no interface provides an unfair advantage. We estimated that for roughly 30\% of comprehension questions, the answers to a discord question in the Annotated Article could be directly useful in finding answers, however, we judge this to be a confirmation of the ability of discord questions to surface salient content, rather than an unfair advantage since the surface question formulations were different.

We studied the answerability of each comprehension question based on the information in each interface. We found that at least one answer element is provided for each question in the Headline List and Annotated Article interfaces, but two of the twenty questions were unanswered in the News Article interface, affecting the completability of the questionnaire in that setting. We discuss question choice further in the Limitations section.

\subsubsection{Study Protocol}

The study had a target duration of 25 minutes, split between an introduction, three reading exercises, and a feedback form. Participants first viewed a timed slideshow presenting the reading exercise and filled out a news reading habits form.

Afterward, participants completed three 6-minute reading exercises. For each exercise, they selected a new story from the five available and were assigned to a random interface, in such a way that each participant completed one exercise with each interface -- the News Article, the Headline List, and the Annotated Article -- in random order.

Finally, participants filled out a completion survey reviewing their experience, both prompting for interface ratings and open-ended feedback.

\subsection{Participants}
\label{sec:nov_participants}

\begin{figure*}
    \centering
    \sbox{\measurebox}{%
      \begin{minipage}[b]{.44\textwidth}
      \subfloat[Agreement to statements about news]
        {\label{fig:nov_survey_state} \includegraphics[width=\textwidth]{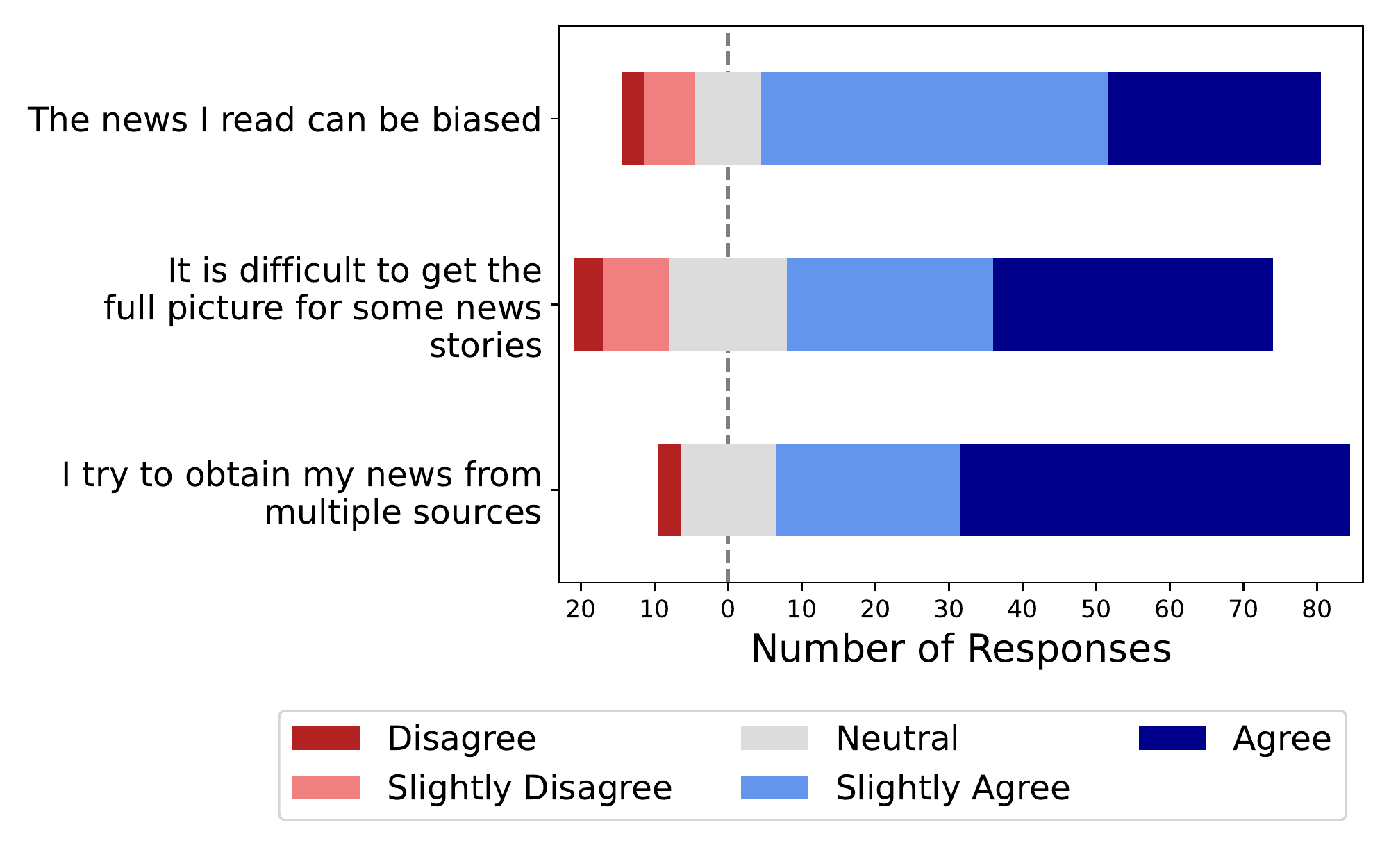}}
      \end{minipage}}
    \usebox{\measurebox}\qquad
    \begin{minipage}[b][\ht\measurebox][s]{.50\textwidth}
    \centering
    \subfloat[Frequency of accessing news]{\label{fig:nov_survey_freq}\includegraphics[width=\textwidth]{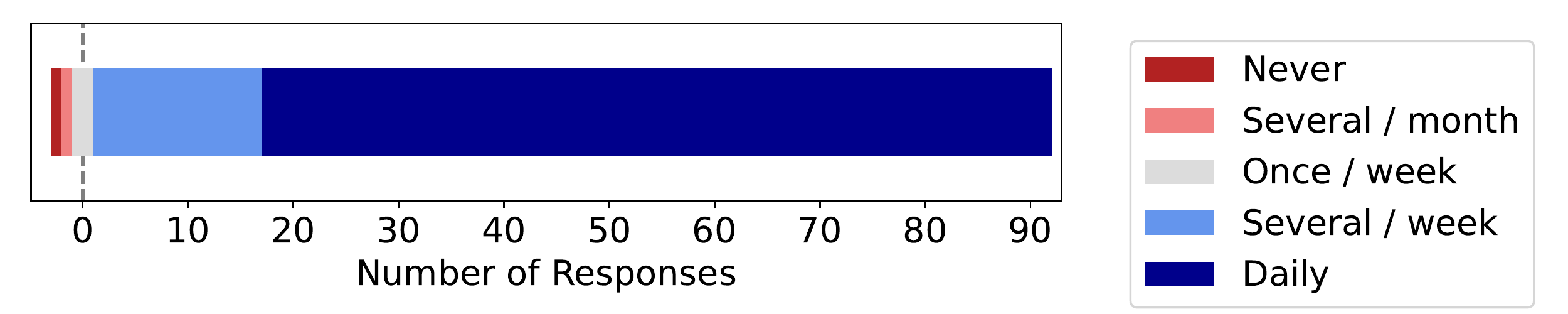} }
    \vfill
    \subfloat[Platform used to access news]{\label{fig:nov_survey_platform} \includegraphics[width=\textwidth]{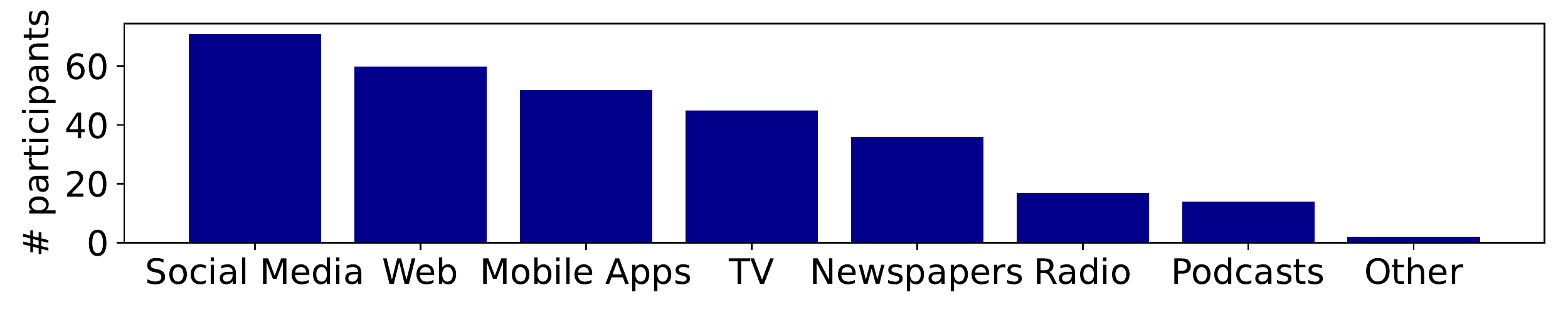}}
    \end{minipage}
    \caption{\textbf{Responses from novice participants on news reading habits.} Participants were asked: (\hyperref[fig:nov_survey_state]{a}) whether they agree with three statements about news consumption, (\hyperref[fig:nov_survey_freq]{b}) how frequently they access the news, and (\hyperref[fig:nov_survey_platform]{c}) which platform they use.}
    \label{fig:novice_survey}
\end{figure*}

Participants were recruited on the Amazon Mechanical Turk crowd-sourcing platform\footnote{https://www.mturk.com}, completing the 25-minute unmoderated task for compensation of \$7 (\$16/hour). We set requirements for recruited participants: be residents of the US, have completed 5000+ tasks on the platform, and have a 98\%+ acceptance rate. Participant results were manually reviewed to further improve result quality.

In total 110 crowd-workers completed our task. Participants were removed from the analysis for one of three reasons: (1) they completed the study in 20 minutes or less (impossible without cheating disabled buttons), or (2) they navigated to other tabs more than three times during the study (we used JavaScript to detect tab switching and warned participants to remain focused on the task), or (3) their feedback hinted at spam participation (e.g., in the open feedback: ``LIFE IS G00D''). Fifteen participants were filtered-out, and we see the added overhead as a cost of ensuring reproducible results. The results presented are based on task completion from 95 participants we believe completed the task to the best of their ability.\enlargethispage{12pt}

News reading habit questionnaire results are summarized in Figure~\ref{fig:novice_survey}. Roughly 80\% of the participants access the news daily, with the five most common platforms being: social media, web news, mobile application, TV, and newspapers. About 85\% of participants at least slightly agreed that the news they read can be biased and that it is difficult to get the complete picture of some news stories. Most participants agreed that they try to obtain their news from multiple sources.

The results confirm that ordinary news readers struggle to obtain broad coverage diversity of some news stories, and put additional effort into reading from multiple sources, justifying the goals of the Assembly prototypes we built.

\subsection{Manual Scoring of Answers}
\label{sec:manual_scores}

Once the study was completed, we manually reviewed the answers. For each question, we extracted all unique answer aspects, considering any aspect that could be a direct answer to the question. For example, eight answer aspects were extracted for the question ``What are the reasons for the Baby Formula shortage?'': (1) the Abbott plant closure, (2) supply chain issues, (3) low retailer stocks, (4) hard-to-find specialized products, (5) a recall due to child sickness, (6) hoarding by some consumers, (7) government inaction, and (8) high prices. Each answer was then scored based on the number of aspects it mentioned. Additional details on the annotation procedure are included in Appendix~\ref{appendix:manual_scoring}. 

We then aggregated the scores based on the reading interface used and computed four metrics: (1) the average score (\textbf{Score}), (2) the percentage of answers left blank intentionally (\textbf{\%No Ans}), (3) the percentage of answers with a score of 0 (\textbf{\%S0}), and (4) the percentage of answers with a score of two or more (\textbf{\%S2+}). We hypothesize that an interface that exposes its user to more coverage diversity will lead to higher average scores and multi-aspect answers (\textbf{Score}, \textbf{\%S2+}), and lower unattempted and zero-score answers (\textbf{\%No Ans}, \textbf{\%S0}).

\subsection{Study Results}
\label{sec:nov_results}

\subsubsection{Quantitative Results}
\begin{table*}[]
    \begin{tabular}{lcccccccc}
     & \multicolumn{4}{c}{\textbf{Answer Scores}} & \multicolumn{4}{c}{\textbf{Statistics}} \\
    \cmidrule(r){1-1} \cmidrule(r){2-5} \cmidrule(r){6-9}
    \textbf{Interface} & \textbf{Score $\uparrow$} & \textbf{\%No Ans $\downarrow$} & \textbf{\%S0 $\downarrow$} & \textbf{\%S2+ $\uparrow$} & \textbf{\#Links} & \textbf{\%Any L} & \textbf{\#Words $\downarrow$} & \textbf{\#Min $\downarrow$} \\
    \cmidrule(r){1-1} \cmidrule(r){2-5} \cmidrule(r){6-9}
    News Article $\clubsuit$ & 0.96\hspace{2.5ex} & 22.1 \hspace{2.5ex} & 36.5 \hspace{2.5ex} & 22.9 \hspace{2.5ex} & -- & -- & 457 & \textbf{6m13s}\aptLtoX[graphic=no,type=html]{\&\#x2662;}{$\diamondsuit$}\aptLtoX[graphic=no,type=html]{\&\#x2661;}{$\heartsuit$} \\
    Headline List \aptLtoX[graphic=no,type=html]{\&\#x2662;}{$\diamondsuit$} & 1.37 $\clubsuit$\hspace{1.0ex} & 6.0 $\clubsuit$ & 20.3 $\clubsuit$\hspace{1.2ex} & 39.4 $\clubsuit$\hspace{1.2ex} & 4.3 & 83.1 & 2490 & 8m46s \\
    Annotated Article \aptLtoX[graphic=no,type=html]{\&\#x2661;}{$\heartsuit$} & \textbf{1.61} $\clubsuit$\aptLtoX[graphic=no,type=html]{\&\#x2662;}{$\diamondsuit$} & 7.6 $\clubsuit$ & \textbf{15.0} $\clubsuit$\aptLtoX[graphic=no,type=html]{\&\#x2662;}{$\diamondsuit$} & \textbf{48.4} $\clubsuit$\aptLtoX[graphic=no,type=html]{\&\#x2662;}{$\diamondsuit$} & 1.3 & 42.1 & 1570 & 7m05s \\
    \cmidrule(r){1-1} \cmidrule(r){2-5} \cmidrule(r){6-9}
    Upper-Bound & 7.58 & -- & -- & 100 & 46.8 & -- & 21100 & --  \\
    \cmidrule(r){1-1} \cmidrule(r){2-5} \cmidrule(r){6-9}
    \end{tabular}
    \caption[]{\textbf{Quantitative results of the reading exercise.} Participants used three interfaces: News article, Headline List, and Annotated Article. Each answer was manually assigned a \textbf{score} based on the number of answer aspects it provides, \textbf{\%No Ans}, \textbf{\%S0}, and \textbf{\%S2+} are the percentage of non-attempted answers, answers with a score of 0, and answers with a score of two or more. \textbf{\#Links}, \textbf{\%Any L}, correspond to the number of links and percentage of exercises where 1+ link was opened. \textbf{\#Words} is the number of words shown to the users, and \textbf{\#Min} average time spent. Column arrows indicate whether higher ($\uparrow$) or lower ($\downarrow$) values are better. We provide upper-bound values for certain elements. Each interface is assigned a symbol ($\clubsuit$\aptLtoX[graphic=no,type=html]{\&\#x2662;}{$\diamondsuit$}\aptLtoX[graphic=no,type=html]{\&\#x2661;}{$\heartsuit$}), which is used to signify pair-wise statistical significant difference ($p<0.05$).}
    \label{table:novice_main}
\end{table*}

Table~\ref{table:novice_main} summarizes the main quantitative results, based on the manual score analysis as well as statistics collected during the study. In terms of statistics, we measured how frequently (\textbf{\%Any L}) and how many (\textbf{\#Links}) links users opened during each exercise (apart from the News Article which does not provide external links) and used link openings to extrapolate how many words were presented to the user (\textbf{\#Words}). We could not measure how many words users actually read, as we did not use eye-tracking technology. To compare quantitative results across settings, we use a t-test to measure pairwise statistical differences.

We find that exercises completed with the Annotated Article lead to the highest overall comprehension scores (mean 1.61), statistically higher than the Headline List (mean 1.37) and the News Article (mean 0.96), confirming that discord question annotations succeed at exposing users to coverage diversity in a realistic reading scenario.

The percentage of questions left blank is much larger for the News Article interface (22.1\%) than the other two (6-8\%), reflecting the lack of exposure from reading a single news article. News Article users' answers receive a score of zero 36.5\% of the time, more than twice as much as the users of the Annotated Article (15.0\%), showing that annotations in the article help participants locate more answer elements and answer a larger fraction of the questions.

The Annotated Article leads participants to produce significantly more multi-aspect answers (\textbf{\%S2+}), with almost half the answers achieving such a score, compared to 39\% for the Headline List and 22.9\% for the news article.

Headline List users predominantly needed to open external links (83\% of the time) showing that simply reading headlines is not sufficient to answer news reading comprehension questions. Because Headline List users opened an average of 4.3 news articles, they were expected to read or skim through much more content (on average 2,490 words), which would take 16 minutes to read at a pace of 150 words per minute. This content overload explains why Headline List participants remained in the interface longer than the two others, staying more than two additional minutes to complete the exercise compared to the delimited time.

In comparison, Annotated Article users completed in roughly 7 minutes, and opened an external link 42\% of the time, with an of average 1.3 links in each session. This is noteworthy, as Annotated Article users were able to achieve higher scores while viewing less textual content (1,570 words on average). This finding underlines the efficiency of discord questions in highlighting coverage diversity, which allowed participants to understand more with less effort.

News Article users were only given a single news article with no external links, and they completed the task mostly on time at the cost of low scores, leaving 22\% of the questions blank. However, they were still able to provide multi-aspect answers for 22.9\% of questions, providing evidence that single news articles occasionally provide multi-perspective news coverage. The quality of the chosen news article has a direct effect on performance, and this study attempts to establish average performance by selecting a news article of median length. Future work can expand on this by studying the extremes as well, the average score received using the longest and shortest news article in the collection.

When looking at upper-bound values, the stories in the study had an average of 46.8 sources, totaling on average 21,100 words per story, which would take more than eleven hours to read in its entirety (at a rate of 150 wpm) for the five stories, underscoring the impossibility of exhaustive reading. The maximum achievable score (7.58) largely surpasses average interface scores (0.9-1.6) highlighting that the designs we propose are the first steps in improving access to news coverage diversity, and there is a large room for improvement.

\subsubsection{From Exposure To Persuasion}
\label{section:persuasion_results}
\begin{table}[]
    \begin{tabular}{lccc}
    \cmidrule(r){1-1} \cmidrule(r){2-4}
    \textbf{Interface} & \textbf{\%1-side $\downarrow$} & \textbf{\%Hypo $\uparrow$} & \textbf{\%2-Side $\uparrow$} \\
    \cmidrule(r){1-1} \cmidrule(r){2-4}
    News Article $\clubsuit$ & 26.7 & 53.3 & 20.0\hspace{3.5ex} \\
    Headline List \aptLtoX[graphic=no,type=html]{\&\#x2662;}{$\diamondsuit$} & 34.6 & 46.2 & 19.2\hspace{3.5ex} \\
    Annotated Article \aptLtoX[graphic=no,type=html]{\&\#x2661;}{$\heartsuit$} & 14.3 & 23.8 & 61.9 $\clubsuit$\aptLtoX[graphic=no,type=html]{\&\#x2662;}{$\diamondsuit$} \\
    \cmidrule(r){1-1} \cmidrule(r){2-4}
    \end{tabular}
    \caption[]{\textbf{Breakdown of answer categories for prediction question.} When asked to predict the future likelihood of a recession, participants could answer with a one-sided assurance (\textbf{\%1-side}), with a hypothetical (\textbf{\%Hypo}), or with a two-sided answer (\textbf{\%2-Side}). Each interface is assigned a symbol ($\clubsuit$\aptLtoX[graphic=no,type=html]{\&\#x2662;}{$\diamondsuit$}\aptLtoX[graphic=no,type=html]{\&\#x2661;}{$\heartsuit$}), which is used to signify pair-wise statistical significant difference ($p<0.05$).}
    \label{table:nov_persuasion}
\end{table}

Overall, the exercise we design focuses on exposure and does not evaluate persuasiveness. Out of the twenty comprehension questions, one of them asked participants to make a prediction about future events, allowing us to investigate participant perception of future events. More specifically, in the ``Potential Recession'' exercise the third comprehension question asked for a prediction: ``Is there going to be a recession?''.  We isolate answers to this question for detailed inspection.

In total, 62 participants selected this story. By manual analysis, answers were assigned to three categories: (1) \textbf{one-sided}, when participants expressed certainty on the outcome (e.g., \textit{We are already in it!}, P31), (2) \textbf{hypothetical} when participants express some level of uncertainty (e.g., \textit{Likely. Might be avoided after consumers return to normal spending patterns}, P47), or (3) \textbf{two-sided} when participants express two opinions or more (e.g., \textit{Yes, if recession proves to be the only way to get inflation under control. No, if the Fed can engineer a ``soft landing,''}, P9).

We argue that a successful interface should accompany readers in discovering the uncertainty, reducing the proportion of one-sided answers, and increasing two-sided and hypothetical answers. Results are summarized in Table~\ref{table:nov_persuasion}.

Only 14.3\% of Annotated Article users gave one-sided answers, compared to 26-34\% of baseline interfaces, but the difference is not statistically significant. However, an increase in two-sided answers with the Annotated Article is statistically significant, showing that the Annotated Article encouraged participants to consider several potential alternatives. This result is limited yet encouraging, and we encourage future related work to increase the proportion of open-ended prediction questions, as they are a useful tool in measuring participant persuasion.

\subsubsection{Completion Questionnaire}

\begin{figure*}
    \centering
    \hfill
    \begin{subfigure}[b]{0.384\textwidth}
        \includegraphics[width=\textwidth]{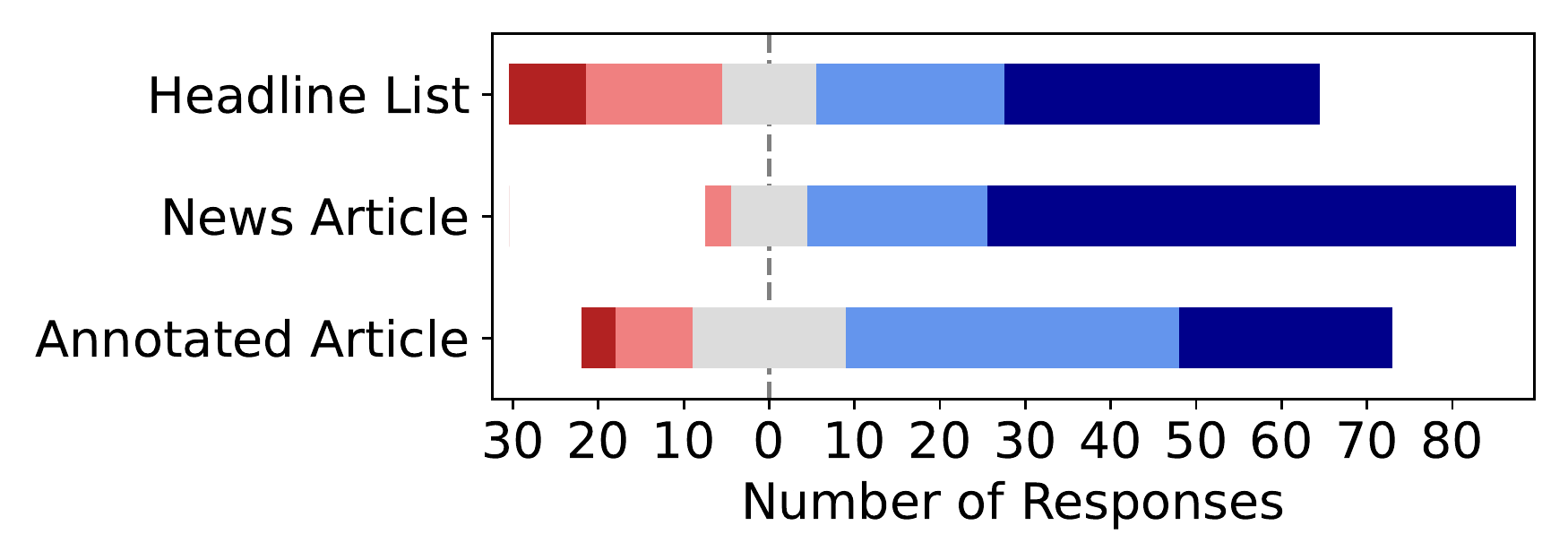}
        \caption[]{Is \textbf{easy to use}}
        \label{fig:nov_easy_expert}
    \end{subfigure}
    \begin{subfigure}[b]{0.50\textwidth}
        \includegraphics[width=0.81\textwidth]{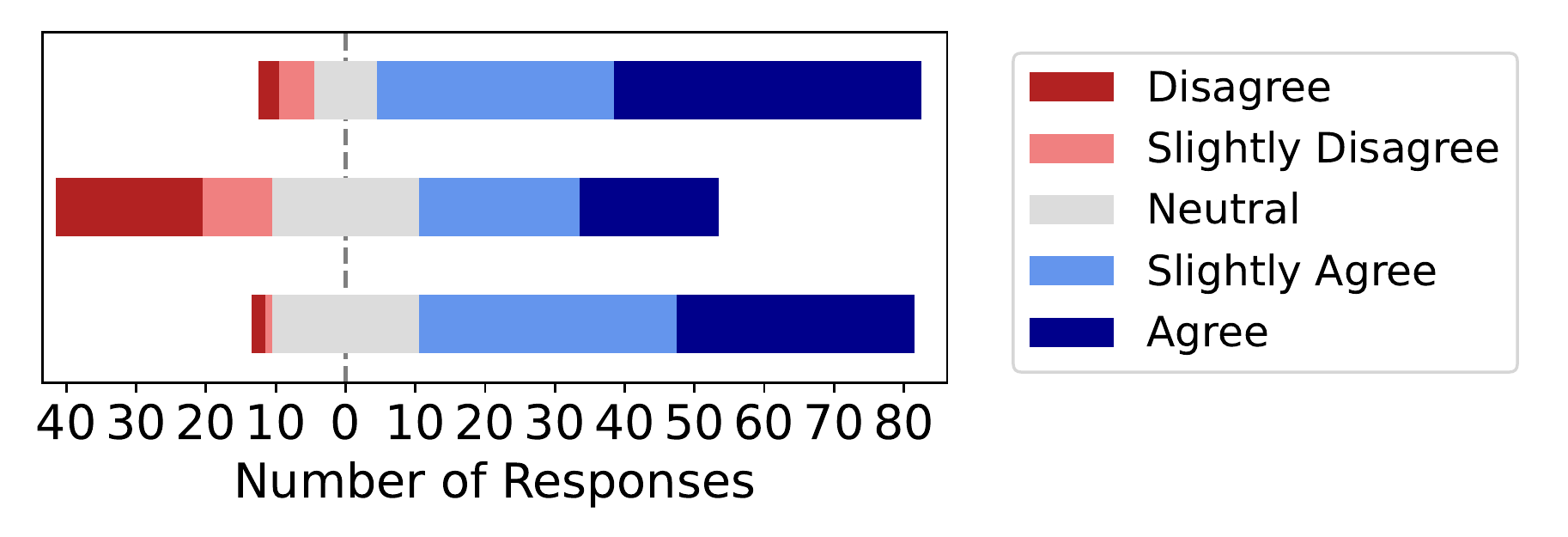}
        \caption[]{Highlights \textbf{coverage diversity}}
        \label{fig:nov_easy_novice}
    \end{subfigure}
    \hfill
    \caption[]{\textbf{Results of Novice Completion Questionnaires.} Upon completing the three reading exercises, participants were asked to rate each interface on whether it is easy to use, and if it highlights coverage diversity (a term definition was provided).}
    \label{fig:novice_likers}
\end{figure*}

In the completion questionnaire, participants were asked to rate each interface on two features: whether it is easy to use, and whether it highlights coverage diversity. We labeled each interface with the name of the story they read in it, to avoid interface name confusion.

Regarding ease of use, the News Article was largely preferred, followed by the Annotated Article and finally the Headline List. The pronounced gap was not predicted by experts, which had estimated all three interfaces to be usable by novice users. We hypothesize this gap is due to the difference in study objective: experts browsed the interface freely, while novice users were completing a comprehension exercise, which might affect impressions of ease of use.

Regarding coverage diversity highlighting, the News Article obtained the lowest ratings, and the Headline List and Annotated Article were virtually tied. Surprisingly, although Annotated Article users scored significantly higher on the comprehension questions, they did not rate the interface higher in terms of coverage diversity. This shows the difficulty for users of news-reading interfaces to directly assess coverage diversity exposure, similar to findings from prior work \cite{heitz2022benefits}, and the usefulness of comprehension questionnaires in measuring user exposure.

\subsection{Qualitative Feedback}
\label{sec:nov_qualitative}

In their open-ended feedback, participants were asked to reflect on: (1) their favorite and least favorite interface, and (2) any aspect that did or did not work well. We employed a thematic analysis \cite{braun2012thematic} with the feedback, grouping by the interface, and filtering to themes mentioned by at least five participants.

\subsubsection{News Article} The News Article was the favorite interface of 10 participants, and the least favorite of 35. Twelve participants mentioned ease of use as the main benefit (e.g., \textit{``Of course the single story is very easy to read''} - P66). The central disadvantage discussed by 30 participants is insufficient information preventing participants from answering all comprehension questions (e.g., \textit{The basic news article was my least favorite as it did not provide enough information.} -- P8).

\subsubsection{Headline List} The Headline List was mentioned as a favorite by 20 participants, and least favorite by 20 participants, showing a polarity among participants. Two positive features stood out, first, 13 participants thought the interface gave access to the most information (e.g., \textit{because I was able to see so many more options and 'sides' to the story} -- P26), and second, seven appreciated doing their ``own research'' (e.g., \textit{The headline list was more like a traditional search engine result page and so it was the most known interface.} -- P71).

Negative features were more diverse: 16 participants complained of a lack of time (e.g., \textit{it would be the most useful with a lot more time to get used to it} -- P5), 15 struggled with opening articles that were later behind a paywall, deleted, or with too many ads (e.g., \textit{but there were so many with subscription blocks and ad-littered articles.} -- P59), 11 disliked the back-and-forth required between picking a headline and reading the full article (e.g., \textit{The headline list was way too broad [...] since you had to go to the articles and read them to see if they even answered questions.} -- P67), 7 did not like the sources listed (e.g., \textit{The headline list seemed to only be 'western' or 'western' friendly perspectives, nothing from China media/propaganda.} - P89), and 7 expressed distrust of headlines due to their potential click-bait and vague nature (e.g., \textit{It is harder to sift through it all with just headlines which in my experiese[sic] are sometimes misleading.} -- P41).

\subsubsection{Annotated Article} The Annotated Article was the favorite interface of 35 participants, and the least favorite of 12. There were three negative aspects discussed: 12 participants found the layout of the interface and particularly the annotations challenging to navigate (e.g., \textit{but the dropdown lists were sort of jarring} -- P84), 11 participants found the annotations could provide incomplete information (e.g., \textit{and wasn't always the most accurate in the information it provided.} - P71), and 5 participants found that annotations were not placed in a pertinent position (e.g., \textit{[did not] connect in an organic way to their placement within the larger article} -- P9).

In terms of positive features, 17 participants found the positioning of annotations to be adequate, sometimes coming as reflections occurred while reading the article (e.g., ``Gave the story and added links to how you can learn more information without overtaking the article.'' -- P86).

\subsection{Summary of Results}

In conclusion, the execution of the reading exercise with 95 participants to compare three interfaces -- News Article, Headline List, and Annotated Article -- reveals significant differences in access to coverage diversity between interfaces. Annotated Article users are more successful at answering comprehension questions with multiple aspects while finding the interface slightly easier to use than the Headline List.

The Annotated Article is however still more challenging to use than an (unannotated) news article, with imperfections in the automatically generated annotations and dropdown design of the annotations causing some user dissatisfaction.

The Headline List falls between the other two interfaces, exposing users to more diversity than a single News Article but less than the Annotated Article. The interface received more polarized feedback, with some users liking the manual aspect of doing their own research, while others were frustrated with having to go back and forth due to uninformative news articles. We hypothesize that the Headline List is most adequate in longer reading sessions when a user has ample time to navigate between articles (e.g., 1 hour to do research on a topic), but less adapted to shorter reading sessions, explaining the frustration of some of its users in our 6-minute exercises.

In Appendix~\ref{appendix:reproducibility}, we assess the reproducibility of the study design through bootstrap re-sampling \cite{efron1982jackknife} and find that statistical significance mostly holds when varying the selection of stories and that result reliability requires recruiting at least 60 participants. 

\section{Discussion \& Limitations}
\label{sec:limitations}

\textbf{Imperfect NLP Methodology.} By choosing to work with an existing NLP framework, we evaluate automatically generated interfaces, measuring whether the value added by discord questions outweighs the noise introduced by imperfect NLP. For example, in Figure~\ref{fig:annotated_article}, the framework extracted the answer ``Peaked'' to the question ``Who does inflation affect?'', which is invalid. The findings are therefore tied to model quality and are likely to change as NLP methods mature. We considered manually post-editing discord questions in the study interfaces to obtain feedback on an ideal version of the interfaces. Still, we preferred the realistic scenario of the fully automatic interface. As suggested by experts in \S\ref{sec:experts_study}, post-editing in a newsroom could ensure the quality of discord questions for production settings.

\textbf{Amazon Mechanical Turk population bias.} Even though recruiting through a crowd-sourcing platform typically leads to more representative participants than on-site recruitment (typically, undergraduate students at a university) \cite{Berinsky2012EvaluatingOL}, the crowd-worker population differs from the U.S. population on many metrics (e.g., age, gender, income level) \cite{Ipeirotis2010DemographicsOM} which would likely have an impact on the results of our second usability study. All participants also received a monetary reward for completing the study, which could affect the authenticity of the interactions we base our results on. We aim to release a public version of prototyped interfaces, which would allow us to observe users interacting with our tools in a more genuine setting.

\textbf{Completability of the Comprehension Questionnaire.} We relied on experts to select comprehension questions to evaluate news readers on. In order to avoid biasing the question towards information in one interface, in particular, the experts were given access to all the content of the news event and would propose comprehension questions irrespective of the studied interfaces. Because of this choice, some of the selected questions are not answerable for some interface-story combinations, reducing the completability of the comprehension questionnaire in some settings. We believe that enforcing the answerability of the questions would be detrimental, as it would lower the difficulty of the comprehension questions which could artificially favor simpler interfaces. We did not however analyze the effect of unanswerability on our results.\enlargethispage{-12pt}

\textbf{Realistic Reading Setting.} The participants in the reading exercise were instructed to complete the comprehension questionnaire while making use of the reading interface in a limited time, making for a more active and focused reading setting which is not representative of more passive news reading. Some participants noted in their feedback that the exercise felt like a research assignment rather than news reading. It is possible that the effect of the discord questions would vary in a more passive news reading scenario, and future work could focus on passive reading scenarios.

\textbf{Recomposed Article Limitations.} The Recomposed Article design can be seen as a middle-ground between the more straightforward Annotated Article and the information-dense Question Grid. The interface received the least enthusiastic feedback from interviewed experts, which found it more confusing than the Annotated Article, while not offering as many analysis opportunities as the Question Grid. We hypothesize that the sorting algorithm in Figure~\ref{fig:algo_recomposed_article} used to compose the article is overly simplistic and leads to a lack of story coherence. The idea of composing novel articles that efficiently present points of discord remains promising, and perhaps generative question-answering models \cite{tafjord2021general,kwiatkowski2019natural} will offer new avenues for improvements in this domain.

\textbf{Headline List Order.} Some participants found the order of headlines in the interface to lack meaning. We simply reproduced the order in which headlines appeared on Google News' original page, and did not further deduplicate or re-order headlines. Prior work could be integrated to organize the headlines into groups \cite{laban2021news,bambrick2020nstm}, or provide information on relations between headlines\cite{gusev2021headlinecause}.

\textbf{Discord Questions for non-news data.} In this paper, we focus the interface design and usability study on the news domain, and interview journalists with expertise in news production. However, the Discord Questions framework \cite{laban2022discord} is not restricted to the news domain. It could prove useful in other multi-document exploration settings, such as helping shoppers navigate 100+ reviews of a product, or instructors navigate the end-of-course student feedback. Although some design components in our work might transfer to new domains, others require domain-specific adaptations.

\textbf{English-Only Data Source.} Our current prototype is inherently limited due to our focus on English-written news sources, as coverage diversity on international topics is likely to come from non-English news sources \cite{maier2020world}. However, improvements in automatic news translation \cite{tran2021facebook}, as well as multi-lingual models \cite{hu2020xtreme} draw a path towards a multi-lingual version of our prototype.

\textbf{Realizability of Advanced Use Cases.} Several of the interviewed experts were enthusiastic about the Question Grid, proposing several scenarios within a newsroom that could benefit from a Question Grid interface. In this paper, we did not follow up with these ideas, focusing instead on evaluating novice-compatible interfaces. Future work can however collaborate with members of newsrooms to tailor a Question Grid-like interface to journalism applications.\enlargethispage{-12pt}

\section{Future Work}

\textbf{Long-Term Effects.} Participants in our studies spent at most 20 minutes with a given interface: enough time for an initial opinion, but not enough for prolonged use judgment. Repeated access to diverse coverage for an ongoing story might have complex effects on readers, either gradually assisting them in diversifying their understanding, or causing them to distrust certain sources over time \cite{koch2013helpful}. Extending the reading exercise to a longitudinal study could provide insights into the interface's long-term effects, but would likely prove challenging to implement. Prior work has built interfaces specific to long-ranging news stories\cite{shahaf2010connecting,laban2017newslens}, and their adaptation with the Discord Question framework is a promising research direction.

\textbf{Design and Deployment Recommendations.} The findings from our usability can provide insights to designers of future news-reading interfaces eyeing to facilitate access to diversity in news coverage. First, questions -- generated or manually curated -- can serve as a tool for alignment of source stances and can trigger curiosity in the reader. Second, as the user reads a single source content, the in-context positioning of the additional source's opinion or analysis (such as in the Annotated Article) can effectively indicate to the user which parts of the news story are complex, and lower the barrier to access broad coverage for curious readers. Finally, in multi-source presentations, transparently indicating the origin of each content element, and the use of hyperlinking to allow access to original source presentation is crucial to maintain user trust.

\textbf{Reproducibility of Results.} We designed the reading exercise with the objective to maximize reproducibility and minimize potential bias toward any of the conditions in the study, for example by recruiting external experts to select the comprehension questions, or through an anonymized scoring of participants' answers (see Appendix~\ref{appendix:manual_scoring}). Statistical testing described in Appendix~\ref{appendix:reproducibility} confirm that the study design should be robust to a different selection of news stories, and a different group of participants, as long as 60 participants or more are recruited. We hope future work can use our study design as a template to measure the efficacy of reading interfaces at surfacing coverage diversity. Some adaptations will likely be necessary, such as selecting newer, more relevant news stories, or updating the selection of NLP models to the latest generation.

\textbf{Measuring Reader Persuasion.} We did not explicitly plan to measure user persuasion at the inception of the project and were focused on reader exposure to content diversity. However, the selection by experts of a prediction question in one of the story's comprehension questionnaires offered an opportunity for the analysis in Section~\ref{section:persuasion_results}, revealing minor evidence that participants might have been influenced in their perception of the likelihood of a future event. We encourage future work to use prediction questions in future usability studies to expand on our preliminary findings.

\section{Conclusion}

This paper introduced three news reading interfaces -- the Annotated Article, the Recomposed Article, and the Question Grid -- that leverage the Discord Questions automated pipeline to add context to news stories and highlight coverage diversity. We first conducted a usability study with journalism experts, gaining insights into potential use cases of the different interfaces. Experts find that discord questions help highlight coverage diversity in all Assembly interfaces, and judge that the Annotated Article is generally accessible to a wide audience, while the other two are suitable for advanced use within newsrooms. In a second study with 95 novice news readers, we assess the usability of the Annotated Article compared to existing news interfaces -- a single News Article, and a Headline List. Novice readers found the Annotated Article roughly as easy to use as existing interfaces while scoring significantly higher on a story understanding questionnaire that measured user exposure to coverage diversity. The findings demonstrate that NLP technology can be integrated into news reading interfaces to assist readers in gaining diverse views on complex news stories, reducing the barrier to educating informed citizens.

\begin{acks}
We thank Jesse Vig, Marti Hearst, Alex Fabbri, and the CHI reviewers for their helpful feedback on the manuscript.
\end{acks}

\bibliographystyle{ACM-Reference-Format}
\bibliography{biblio}

\newpage
\appendix

\section{Recomposed Article Content Algorithm}

Figure~\ref{fig:algo_recomposed_article} introduces a pseudo-code version of the algorithm used to sort through and select the discord questions that are presented in the Recomposed Article.

\begin{figure*}
    \centering
    \begin{python}
def select_discord_qs(questions):
    # Initialize output and progress word count
    selected_qs, seen_paragraphs = [], set([])
    while True:
        # Score each discord question
        for q in questions:
            # A question's score is the number of unseen answering paragraphs
            q.score = len(q.paragraph_set - seen_paragraphs)
        # Select the highest_scoring question
        selected_q, score = questions.select_best_q()
        if score == 0:
            # Exit if no question introduces unseen paragraphs
            break
        # Add question to final list
        selected_qs.append(selected_q)
        # Mark question's paragraph as seen
        seen_paragraphs = seen_paragraphs | selected_q.paragraph_set
    return selected_qs
    \end{python}
    \caption{Python code for the Composition Algorithm used for sequence selection in the Recomposed Article interface.}
    \label{fig:algo_recomposed_article}
\end{figure*}

\section{Baseline Interfaces}
\label{appendix:baseline_interfaces}

\begin{figure*}
    \centering
    \fbox{
    \includegraphics[width=0.7\textwidth] {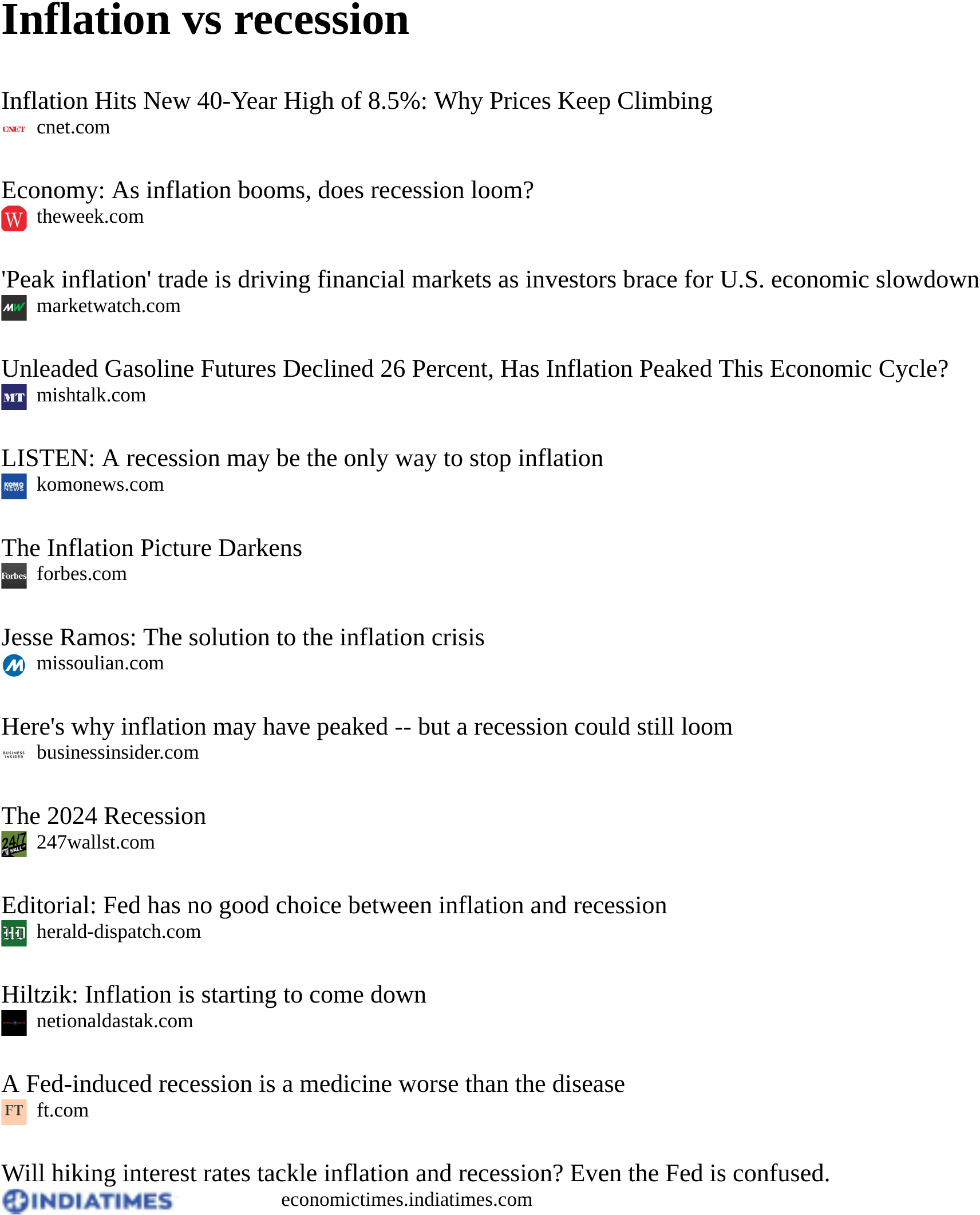}
    }
    \caption{\textbf{Headline List Interface}. We purposefully imitate the style of the ``More Coverage'' view in Google News, listing all the source's headlines within a news story.}
    \label{fig:headline_list_interface}
\end{figure*}

Figure~\ref{fig:headline_list_interface} presents the Headline List baseline interface, which is simply a list of the source's headline. Each headline is clickable and opens the original website in a new browser tab.

\section{Reading Exercise Interface}
\label{appendix:exercise_interface}

\begin{figure*}
    \centering
    \fbox{
    \includegraphics[width=0.7\textwidth]{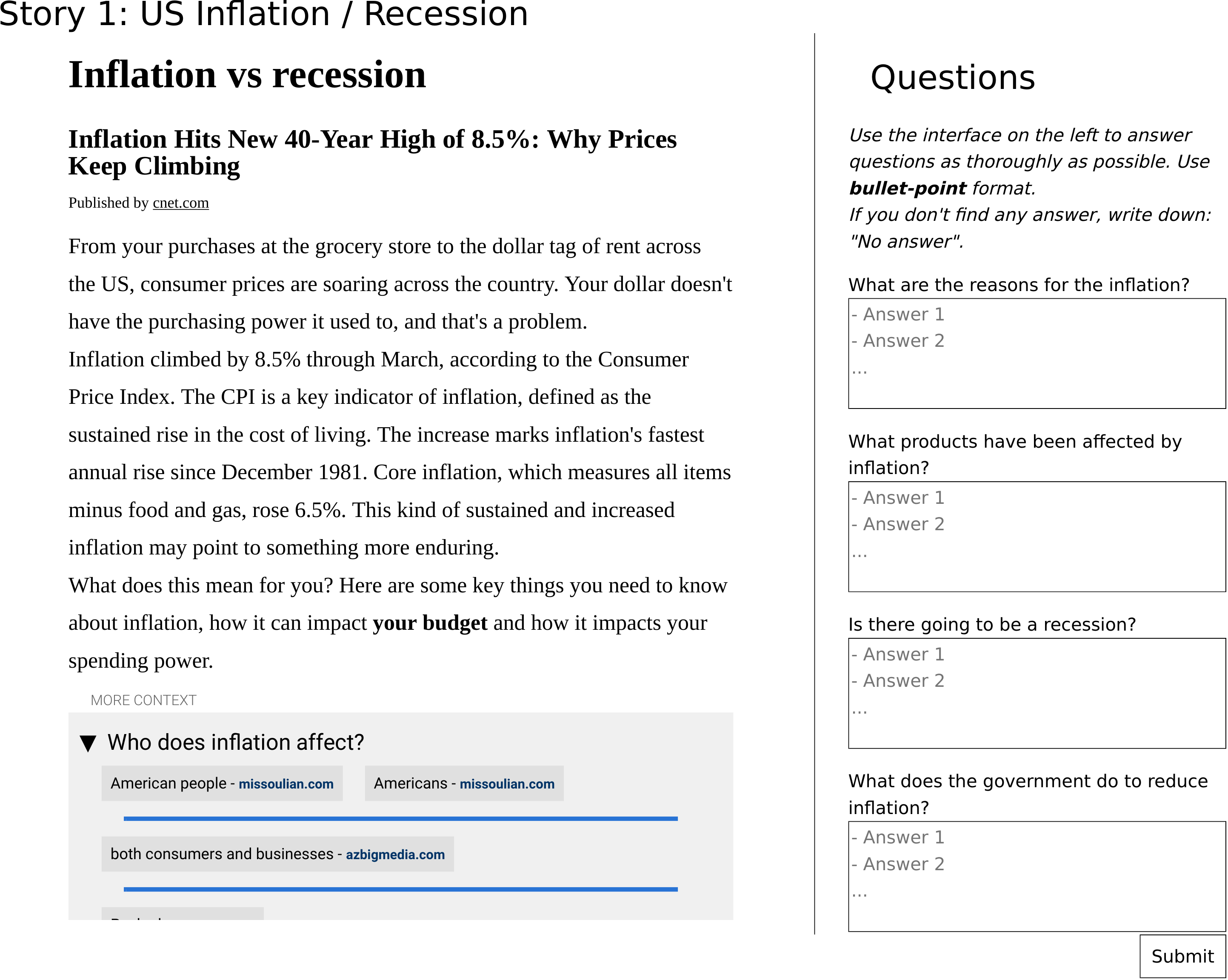}
    }
    \caption{\textbf{Two-column interface used during reading exercise}. Participants completed the comprehension questions in the right column while using the reading interface in the left column. In this case, the exercise was assigned with the Annotated Article reading interface.}
    \label{fig:exercise_interface_columns}
\end{figure*}

Figure~\ref{fig:exercise_interface_columns} displays the interface used during the reading exercise study, using a two-column layout for concurrent reading and question answering.

\section{Manual Scoring Procedure}
\label{appendix:manual_scoring}

The scoring of answers was performed manually by the authors of the paper, and we took steps to ensure the impartiality of the scoring. All answers to a question were loaded into a Google Sheet, with an anonymized identifier that could not readily identify which user wrote each answer, or which interface was used.

In the first step, one reader read through all the answers and identified the set of all answer elements. We required that an answer element to be a plausible answer to the question be added as one possible answer element to be considered in the score. In some cases, participants provided answers that were unrelated to the question, or could not be interpreted as a direct answer, and these were discarded. We did not attempt to assess the veracity of each answer element, or whether each answer was explicitly stated in at least one of the provided articles, as some of the questions required the user's interpretation. For example, to the question ``Is there going to be a recession?'', some participants answered with the answer element: ``we are already in one'', which was factually incorrect during the study dates (according to the definition of a recession requiring two consecutive quarters of economic contraction). In order to maximize our assessment of access to coverage diversity, any plausible answer element was added to the answer element set.

In the second step, the same reader shuffled all the answer elements and tagged each answer with all the answer elements it mentioned. For all answers, we generously assigned the presence of an answer element, for example in cases when the element is partially mentioned or strongly implied.

Once the tagging was completed for all answers, answers were then processed automatically and scores were aggregated. Although we did not perform inter-annotator agreement evaluation for the manual scoring process, and it is likely that there exists some variance in the methodology we used, we believe the process did not favor any particular interface, and this variance should not negatively impact our results or their interpretation.

\section{Study Reproducibility Experiments}
\label{appendix:reproducibility}
\begin{figure*}
    \centering
    \includegraphics[width=0.5\textwidth]{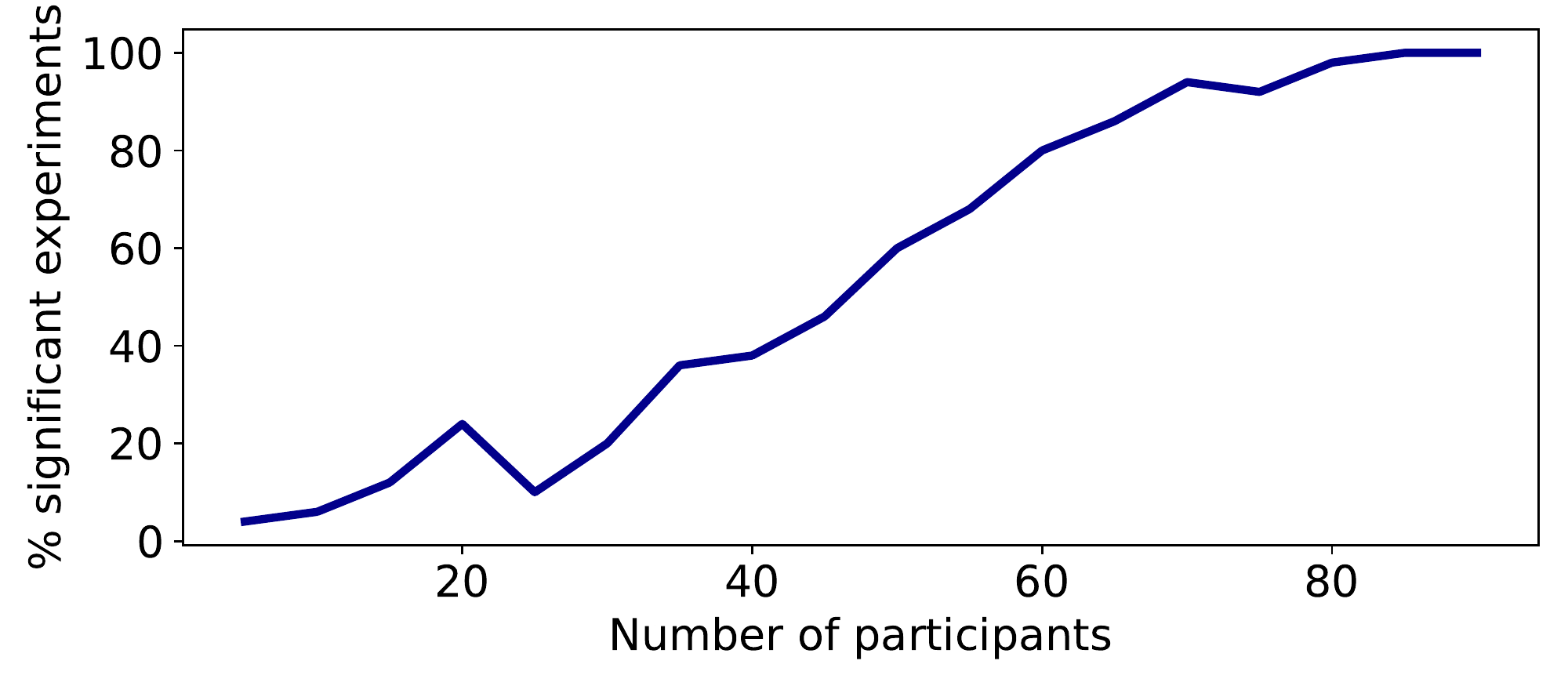}
    \caption{\textbf{Reproducibility experiments for the subset of participants.} As the number of participants decreases, the results are less likely to be significant, showing the benefit of increasing the population size when running the study.}
    \label{fig:reproducibility}
\end{figure*}

In order to determine the level of reproducibility of the reading exercise we designed, we perform two reproducibility experiments, leveraging bootstrap re-sampling \cite{efron1982jackknife} to simulate a change in study results and verify whether the results remain statistically significant. We experiment with two re-sampling methods to verify whether the results remain significant with a different story selection, or a different participant population.

\textbf{Varying the stories.} We consider subsets of the data where only four of the five stories are kept, recompute the average reading scores under each interface and perform a paired t-test to verify whether results remain significant. In 80\% of cases, the score differences are significant ($p<0.05$), confirming that no individual story is responsible for the result, and giving evidence that the results would extend to other stories. When decreasing subset sizes to only 3 out of the five stories, only 66\% of the cases are significant, showing that results are more stable when using at least 4 distinct stories, as variances amongst individual stories exist.

\textbf{Varying the participants.} We simulate a lower number of participants, testing each value of participants from 5 to 95 in increments of 5. For each number of participants, we sample 40 random sets of participants of that size, compute results for that subset and test whether there are statistically significant differences in answer scores. The results are summarized in Figure~\ref{fig:reproducibility}. We find that the results remain largely significant with 60 participants or more, and then the results become less significant as the number of participants decreases. This can serve as a rule-of-thumb for similar studies, encouraging a participant population of at least 60 participants to increase the likelihood of statistical significance in the results.

\end{document}